\documentclass[seceq]{ptptex}

\usepackage{graphicx,slashbox}
\usepackage{wrapft}

\def\ignore#1{{}}
\def\mybig{\displaystyle \strut }

\def\dd{\partial}
\def\la{\raise.16ex\hbox{$\langle$}\lower.16ex\hbox{}  }
\def\ra{\, \raise.16ex\hbox{$\rangle$}\lower.16ex\hbox{} }
\def\go{\rightarrow}

\def\onehalf{ \hbox{${1\over 2}$} }
\def\half{ {1\over 2} }

\def\eff{{\rm eff}}

\newcommand{\beeq}{\begin{equation}}
\newcommand{\eneq}{\end{equation}}
\newcommand{\beqn}{\begin{eqnarray}}
\newcommand{\eeqn}{\end{eqnarray}}
\newcommand{\under}[1]{$\underline{\hbox{#1}}$}
\def\vphi{\varphi}
\def\myfrac#1#2{{\mybig #1\over \mybig #2}}

\newcommand{\alp}{\alpha}
\newcommand{\bt}{\beta}
\newcommand{\gm}{\gamma}
\newcommand{\Gm}{\Gamma}
\newcommand{\dlt}{\delta}

\newcommand{\vep}{\varepsilon}
\newcommand{\tht}{\theta}

\newcommand{\lmd}{\lambda}
\newcommand{\Lmd}{\Lambda}
\newcommand{\sgm}{\sigma}

\newcommand{\vph}{\varphi}
\newcommand{\omg}{\omega}
\newcommand{\Omg}{\Omega}

\newcommand{\be}{\begin{equation}}
\newcommand{\ee}{\end{equation}}
\newcommand{\bea}{\begin{eqnarray}}
\newcommand{\eea}{\end{eqnarray}}
\newcommand{\eql}{&=&}

\newcommand{\defa}{&\equiv&}

\newcommand{\mtrx}[4]{\begin{pmatrix}#1&#2\\#3&#4\end{pmatrix}}
\newcommand{\vct}[2]{\begin{pmatrix}#1\\#2\end{pmatrix}}

\newcommand{\simgt}{\stackrel{>}{{}_\sim}}

\newcommand{\tl}[1]{\tilde{#1}}
\newcommand{\bdm}[1]{{\mbox{\boldmath $#1$}}}
\newcommand{\tr}{{\rm tr}}

\newcommand{\diag}{{\rm diag}}
\newcommand{\der}{\partial}
\newcommand{\dr}{\!\!d}
\newcommand{\hc}{{\rm h.c.}}
\newcommand{\ie}{{\it i.e.}}

\newcommand{\sgn}{{\rm sgn}}

\newcommand{\brkt}[1]{\left( #1 \right)}
\newcommand{\brc}[1]{\left\{ #1 \right\}}
\newcommand{\sbk}[1]{\left[ #1 \right]}
\newcommand{\abs}[1]{\left| #1 \right|}

\renewcommand{\Im}{{\rm Im}}

\newcommand{\cD}{{\cal D}}

\newcommand{\cL}{{\cal L}}

\newcommand{\cO}{{\cal O}}
\newcommand{\cP}{{\cal P}}
\newcommand{\cQ}{{\cal Q}}

\newcommand{\zp}{z_\pi}
\newcommand{\thw}{\tht_W}
\newcommand{\thH}{\tht_{\rm H}}
\newcommand{\cw}{c_{\rm H}}
\newcommand{\sw}{s_{\rm H}}
\newcommand{\tw}{t_{\rm H}}
\newcommand{\cth}{c_\tht}
\newcommand{\sth}{s_\tht}
\newcommand{\cph}{c_\phi}
\newcommand{\sph}{s_\phi}
\newcommand{\tph}{t_\phi}
\newcommand{\ubl}{U(1)_{\rm B-L}}
\newcommand{\suL}{SU(2)_{\rm L}}
\newcommand{\suR}{SU(2)_{\rm R}}
\newcommand{\uy}{U(1)_Y}
\newcommand{\uem}{U(1)_{\rm EM}}
\newcommand{\mKK}{m_{\rm KK}}
\newcommand{\aL}{a_{\rm L}}
\newcommand{\aR}{a_{\rm R}}
\newcommand{\nL}{3_{\rm L}}
\newcommand{\nR}{3_{\rm R}}
\newcommand{\chL}{\pm_{\rm L}}
\newcommand{\chR}{\pm_{\rm R}}

\newcommand{\aLR}[1]{a_{{\rm L}#1{\rm R}}}
\newcommand{\fqR}[1]{\tl{f}^q_{R,#1}}
\newcommand{\fQR}[1]{\tl{f}^Q_{R,#1}}
\newcommand{\fqL}[1]{\tl{f}^q_{L,#1}}
\newcommand{\fQL}[1]{\tl{f}^Q_{L,#1}}
\newcommand{\hfqR}[1]{\hat{f}^q_{R,#1}}
\newcommand{\hfQR}[1]{\hat{f}^Q_{R,#1}}
\newcommand{\hfqL}[1]{\hat{f}^q_{L,#1}}
\newcommand{\hfQL}[1]{\hat{f}^Q_{L,#1}}

\title{
Anomalous Higgs Couplings 
in the $\bdm{SO(5)\times \ubl}$ Gauge-Higgs Unification 
in Warped Spacetime
}

\author{
Yutaka \textsc{Hosotani}\footnote{
E-mail: hosotani@het.phys.sci.osaka-u.ac.jp} 
and Yutaka \textsc{Sakamura}\footnote{
E-mail: sakamura@het.phys.sci.osaka-u.ac.jp} 
}

\inst{
Department of Physics, Osaka University,
Toyonaka, Osaka 560-0043, Japan
}

\abst{
The gauge couplings $WWZ$, $WWWW$, and $WWZZ$ in 
the gauge-Higgs unification scenario in the Randall-Sundrum warped spacetime
remain almost universal as in the standard model, but substantial deviation
results for the Higgs couplings.
In the $SO(5)\times \ubl$ model, the couplings $WWH$ and $ZZH$
are suppressed by a factor $\cos \thH$ from the values in the 
standard model, while the bare couplings $WWHH$ and $ZZHH$ are suppressed 
by a factor $1 - \frac{2}{3} \sin^2 \thH$.  Here $\thH$ is the
Yang-Mills AB phase (Wilson line phase) along the fifth dimension, 
which characterizes the  electroweak symmetry breaking.  
The suppression can be used to test
the gauge-Higgs unification scenario at LHC and ILC. 
It is also shown that the $WWZ$ coupling in  flat spacetime
deviates from the standard model value at moderate values of $\thH$,
contradicting with  the LEP2 data.
}

\begin{document}

\maketitle

\section{Introduction}

In the previous paper we showed that substantial deviation in the 
Higgs couplings to $W$ and $Z$ bosons from those in the standard
model is expected as a general feature in the gauge-Higgs unification 
model in warped spacetime,\cite{SH1} which can be tested at LHC and ILC
in the coming future.   Further the deviation in the $WWZ$ coupling was shown
to be very small in warped spacetime.
In the present paper we give more thorough  analysis of 
these couplings to strengthen the statements, in addition to give detailed 
account of  the mass spectrum and wave functions 
in the gauge-Higgs sector in the  $SO(5)\times \ubl$ model.

In the gauge-Higgs unification scenario the Higgs field in four dimensions 
is identified with the zero mode of the extra-dimensional component of 
gauge potentials in higher dimensional gauge theory.  As such, the mass and
couplings of the Higgs field are not arbitrary parameters in theory.  They follow
from the gauge principle.  The original proposal by Fairlie and
by Manton to unify the Higgs field in the six-dimensional gauge theory with
$S^2$ as extra dimensions was unsatisfactory 
 as it gives too low Kaluza-Klein energy 
scale and unrealistic couplings and spectrum.\cite{Fairlie1,Manton1} \ 
Shortly after their proposal 
it has been recognized that Wilson line phases, or 
Yang-Mills Aharonov-Bohm phases associated with non-simply connected 
extra dimensions can 
serve as the Higgs field in four dimensions.  These phases, denoted as $\thH$
in the present paper, label classically degenerate vacua.   The value of  $\thH$
is determined at the quantum level.  When the value is nontrivial,
the gauge symmetry is dynamically broken.\cite{YH1,YH2} \ 

The scenario of identifying $\thH$ with a 4D Higgs field, 
which was applied first to GUT and then to  the electroweak interactions, 
has many attractive features.\cite{YH3}\tocite{Gersdorff} \ 
Besides inducing dynamical gauge symmetry breaking, it predicts a finite
mass for the Higgs field, independent of the cutoff scale.\cite{YHscgt2}\tocite{YHfinite} \ 
In the electroweak theory, it can solve the gauge  hierarchy problem.\cite{Lim2} \ 
The dynamically determined value of  $\thH$ depends on the details of 
the theory, particularly in the fermion sector.   Astonishingly many  of the
features in the gauge-Higgs sector such as the mass spectrum and
couplings are determined once the value of $\thH$ is specified. 
In this respect our  analysis is robust.  It is shown below that the 
$WWH$, $ZZH$, $WWHH$ and $ZZHH$ couplings are suppressed 
compared with those in the standard model. 
The predictions obtained for the gauge-Higgs couplings can be tested
at LHC and ILC.  If the deviation from the standard model is observed as 
indicated in the gauge-Higgs unification scenario, then it gives strong
hint for the existence of extra dimensions.
It is also confirmed that
the $WWZ$ coupling remains universal in warped spacetime, but it
becomes smaller in flat spacetime compared with that in the standard
model, thus already contradicting with the LEP2 data on $W$ pair production.
This strongly suggests that the extra-dimensional space is curved
and warped, if it exists.  There seems intimate connection between 
the gauge-Higgs unification scenario and the holography 
in the warped space.\cite{Pomarol2,Agashe2,Gherghetta2,Contino1,GGPR} \ 

The paper is organized as follows.  The $SO(5)\times \ubl$ model is
set up in the next section.  The spectrum and mode functions of gauge
bosons are given in Sec.\ 3, whereas those of fermions are given in 
Sec.\ 4.  Approximate masses and wave  functions of gauge fields,  
Higgs field, and light fermions in four  dimensions are given in a 
simple form in Sec.\ 5.  Gauge couplings and Higgs couplings are
evaluated in Sec.\ 6 to make predictions described above.
Gauge couplings of fermions are briefly discussed in Sec.\ 7 and a summary 
is given in Sec.\ 8.  Useful formulas are collected in appendices.

\section{$SO(5)\times \ubl$ model}

We consider an $SO(5)\times \ubl$ gauge theory in the warped 
five-dimensional spacetime.\cite{Agashe2} \ 
The fifth dimension is compactified on an orbifold~$S^1/Z_2$ 
with a radius $R$. 
We use, throughout the paper, $M,N,\cdots = 0,1,2,3,4$ for 
the 5D curved indices, $A,B,\cdots, = 0,1,2,3,4$ for the 5D 
flat indices in tetrads, and $\mu,\nu,\cdots=0,1,2,3$ 
for 4D flat indices.
\ignore{\footnote{As the background geometry preserves 
4D Poincar\'{e} invariance, the curved 4D indices are not 
discriminated from the flat 4D indices. }}
The background metric is given by~\cite{RS1}
\be
 ds^2 = G_{MN}dx^M dx^N = e^{-2\sgm(y)}\eta_{\mu\nu}dx^\mu dx^\nu
 +dy^2, \label{metric}
\ee
where $\eta_{\mu\nu}=\diag(-1,1,1,1), \sgm(y)=\sgm(y+2\pi R)$, 
and $\sgm(y)\equiv k\abs{y}$ for $\abs{y}\leq \pi R$. 
The cosmological constant in the bulk 5D spacetime is given by 
$\Lmd=-k^2$. 
$(x^\mu,-y)$ and $(x^\mu,y+2\pi R)$ are identified with $(x^\mu,y)$. 
The spacetime is equivalent to the interval 
in the fifth dimension~$y$ with two boundaries at $y=0$ and 
$y=\pi R$, which we refer to as the Planck brane and the TeV brane, 
respectively. 

There are   $SO(5)$ gauge fields~$A_M$ and $\ubl$ gauge field~$B_M$. 
The former are decomposed as 
\be
 A_M = \sum_{I=1}^{10}A^I_M T^I 
 = \sum_{\aL=1}^3 A^{\aL}_M T^{\aL}+\sum_{\aR=1}^3 A^{\aR}_M T^{\aR}
 +\sum_{\hat{a}=1}^4 A^{\hat{a}}_M T^{\hat{a}}, 
\ee
where $T^{\aL,\aR}$ ($\aL,\aR=1,2,3$) and $T^{\hat{a}}$ 
($\hat{a}=1,2,3,4$) are the generators of $SO(4)\sim\suL\times\suR$ 
and $SO(5)/SO(4)$, respectively. 
The spinorial representation of $T^I$ 
is tabulated in (\ref{SO_gen}) in  appendix A. 
As a matter field we introduce a spinor field~$\Psi$ 
in the spinorial representation of $SO(5)$ (\ie, $\bdm{4}$ of $SO(5)$)
in the bulk as an example.  
\ignore{Later spinor fields localized on the  Planck brane 
are also introduced to build a realistic model  of quarks and leptons.  }

The relevant part of the action in the bulk is \cite{GP,Chang}
\bea
 S \eql \int\dr^5x\;\sqrt{-G}\left [
 -\tr\brkt{\frac{1}{2}F^{(A)MN}F^{(A)}_{MN}+\frac{1}{\xi}(f^{(A)}_{\rm gf})^2
 +\cL^{(A)}_{\rm gh}}\right.   \label{action1} \\
 \noalign{\kern 10pt}
 &&\hspace{10mm}
 \left.-\brkt{\frac{1}{4}F^{(B)MN}F^{(B)}_{MN}+\frac{1}{2\xi}(f^{(B)}_{\rm gf})^2
 +\cL^{(B)}_{\rm gh}}
 +i\bar{\Psi}\Gm^N\cD_N\Psi-iM_\Psi\vep\bar{\Psi}{\Psi}\right ], \nonumber
\eea
where $G\equiv\det(G_{MN})$ and $\Gm^N\equiv e_{A}^{\;\;N}\Gm^A$. 
The 5D $\gm$-matrices~$\Gm^A$ are related to the 4D ones~$\gm^\mu$ 
by $\Gm^\mu=\gm^\mu$ and $\Gm^4=\gm_5$ which is the 4D chiral operator.
The gauge-fixing functions~$f^{(A,B)}_{\rm gf}$ are specified in the next section. 
$\cL^{(A,B)}_{\rm gh}$ are the associated ghost Lagrangians, and 
$M_\Psi$ is a bulk (kink) mass parameter. 
Since the operator~$\bar{\Psi}\Psi$ is $Z_2$-odd, 
we need the periodic sign function~$\vep(y)=\sgm^\prime(y)/k$ 
satisfying $\vep(y)=\pm 1$. 
The field strengths and the covariant derivatives are defined by 
\bea
 F^{(A)}_{MN} \defa \der_M A_N-\der_N A_M-ig_A[A_M,A_N] ~, \nonumber\\
\noalign{\kern 5pt}
 F^{(B)}_{MN} \defa \der_M B_N-\der_N B_M, \nonumber\\
\noalign{\kern 5pt}
 \cD_M\Psi \defa \brc{\der_M-\frac{1}{4}\omg_M^{\;\;AB}\Gm_{AB}
 -ig_A A_M-i\frac{g_B}{2} \cQ_{\rm B-L} B_M}\Psi, 
\eea
where $g_A$ ($g_B$) is the 5D gauge coupling for $A_M$ ($B_M$), 
$\cQ_{\rm B-L}$ is a charge of $\ubl$, and
 $\Gm^{AB}\equiv\frac{1}{2}[\Gm^A,\Gm^B]$.  
The spin connection 1-form $\omg^{AB}=\omg_M^{\;\;AB}dx^M$ determined from 
the metric~(\ref{metric}) is 
\be
 \omg^{\nu 4} = -\sgm^\prime e^{-\sgm}dx^\nu, \;\;\;\;\;
 \mbox{other components}=0. 
\ee

The boundary conditions at the fixed points $y_0=0$ and $y_\pi = \pi R$,
which preserve  the orbifold structure,  are   
\bea
\hskip -1cm 
&& \vct{A_\mu}{A_y}(x, y_j-y) = P_j \vct{A_\mu}{-A_y}(x,y_j +y)P_j^{-1} 
 ~~, \nonumber\\
 \noalign{\kern 10pt}
\hskip -1cm 
&& \vct{B_\mu}{B_y}(x,y_j -y) = \vct{B_\mu}{-B_y}(x,y_j+ y) 
 ~~, \nonumber\\
 \noalign{\kern 10pt}
\hskip -1cm 
&& \Psi(x, y_j-y) = \eta_j P_j \gm_5\Psi(x, y_j+ y) ~~, 
 \label{Z2parity}
\eea
where $\eta_j=\pm 1$  ($j=0,\pi$).
$P_j\in SO(5)$ are constant matrices  satisfying $P_j^2=1$. 
In the present paper we take   
\be
 P_0 = P_\pi = \mtrx{1_2}{}{}{-1_2} \label{def_Ps} 
\ee
in the spinorial representation, or equivalently 
$P_0 = P_\pi = \diag(-1,-1,-1,-1,1)$ 
in the vectorial representation. 
Then the gauge symmetry is broken to 
$SO(4)\times\ubl\sim \suL\times\suR\times\ubl$ at both boundaries. 
(The broken generators are $T^{\hat{a}}$; $\hat{a}=1,2,3,4$.)
It is convenient to decompose $\Psi$ as 
\be
 \Psi = \vct{q}{Q},  \label{def_qQ}
\ee
where $q$ and $Q$ belong to 
$(\onehalf, 0)$ and $(0, \onehalf)$ 
of $\suL\times\suR$, respectively. 

Fields with Neumann boundary conditions at both boundaries   
have zero modes when perturbation theory is 
developed around the trivial configuration~$A_M=0$. 
With (\ref{Z2parity}) and (\ref{def_Ps}) there arise zero modes 
for $A^{\hat{a}}_y$ ($\hat{a}=1,2,3,4$).
They are identified with the $\suL$ doublet-Higgs field in the standard model; 
$\Phi\propto (A^{\hat{1}}_y+iA^{\hat{2}}_y,A^{\hat{4}}_y-iA^{\hat{3}}_y)^t$. 
A nonvanishing expectation value of $A^{\hat{4}}_y$ 
 gives rise to a Wilson line phase or Yang-Mills Aharonov-Bohm phase, 
 $\thH  \equiv  g_A\int_0^{2\pi R}dy \,   \la A^{\hat{4}}_y \ra  /  2\sqrt{2} 
 =(g_A/ \sqrt{2}) \int_0^{\pi R}dy \, \la A^{\hat{4}}_y \ra$.
More explicitly, 
\be
\la  A^{\hat{4}}_y \ra 
= \frac{2\sqrt{2}ke^{2ky}}{g_A(e^{2k\pi R}-1)} \, \thH ~~. 
 \label{Ay-thH}
\ee
Although $\thH\neq 0$ gives vanishing field strengths, 
it affects physics at the quantum level. 
The global minimum of the effective potential for $\thH$  
determines the quantum vacuum.\cite{YH1} \ 
The nonvanishing $\thH$  induces dynamical electroweak 
gauge symmetry breaking.

There are residual gauge transformations which maintain 
the boundary condition~(\ref{Z2parity}).\cite{YH2,HM} \ 
A large gauge transformation given by
\beeq
\Omega^{\rm large} (y)= \exp \bigg\{ in\pi ~ \frac{e^{2ky} - 1}
{e^{2k\pi R} - 1} ~ (2\sqrt{2} \, T^{\hat 4} ) \bigg\} 
\label{largeGT1}
\eneq
($0 \le y \le \pi R$, $n$: an integer) shifits   
$\thH$ by  $\thH + 2\pi n$,  which implies that
all physical  quantities are periodic functions of $\thH$.
The large gauge invariance   is vital to guarantee the
finiteness of the Higgs boson mass.\cite{YHscgt2,YHfinite} \ 
The $\thH$-dependent part of the effective potential diverges
without the large gauge invariance.

The even-odd property in (\ref{Z2parity}) does not 
completely fix boundary conditions of the fields. 
If there are no additional dynamics on the two branes, 
fields which are odd under parity at $y=0$ or $\pi R$  obey the Dirichlet
boundary condition (D) so that they  vanish there. 
On the other hand,  fields which are
even under parity obey the Neumann boundary conditions (N).  For gauge fields
the Neumann boundary condition is given by
$dA_\mu/dy=0$ or $d(e^{-2ky} A_y)/dy=0$.  
As a result  of additional dynamics on the branes, however, a field with even 
parity, for instance,  can effectively obey the Dirichelet boundary condition.  
The field develops a cusp-type singularity there due to brane dynamics.
As discussed below, the $SO(4)$ symmetry on the Planck brane is broken
to $\suL \times U(1)_Y$ in this manner.

Let us define new fields $A^{\prime \nR}_M$ and $A^Y_M$ by
\beqn
&&\hskip -1cm
 \vct{A^{\prime \nR}_M}{A^Y_M} =
 \mtrx{\cph}{-\sph}{\sph}{\cph} \vct{A^{\nR}_M}{B_M} ~,  \cr
\noalign{\kern 10pt}
&&\hskip -.5cm
 \cph \equiv \frac{g_A}{\sqrt{g_A^2+g_B^2}} ~~,~~
  \sph \equiv \frac{g_B}{\sqrt{g_A^2+g_B^2}} ~~.  
  \label{def_vph}
\eeqn
$A^{a_{\rm R}}_\mu$ and $B_\mu$ are even under parity, whereas
$A^{a_{\rm R}}_y$ and $B_y$ are odd.  
We suppose that as a result of additional dynamics on the Planck brane
the even fields $A^{1_{\rm R}}_\mu$, $A^{2_{\rm R}}_\mu$, and   
$A^{\prime 3_{\rm R}}_\mu$ obey the Dirichlet (D) boundary condition there.
The boundary conditions for gauge fields are tabulated  in Table~\ref{bd_gauge}. 
It is  confirmed that the boundary conditions in Table~\ref{bd_gauge}
preserve the large gauge invariance, that is, new gauge potentials obtained by
(\ref{largeGT1}) obey the same boundary conditions as the original fields.
We remark that  the Neumann   (N) boundary condition on the Planck brane 
 cannot be imposed on $A^{1_{\rm R}}_y$, $A^{2_{\rm R}}_y$, and   
$A^{\prime 3_{\rm R}}_y$, as it does not preserve the large gauge invariance.
Indeed, under a large gauge transformation (\ref{largeGT1}), 
$A_y \go A_y'$ and 
\beeq
\frac{d}{dy} \Big( e^{-2ky} A^{\prime \,\hat a}_y \Big) 
= \frac{d}{dy} \Big( e^{-2ky} A^{\hat a}_y \Big) 
+ \frac{2\sqrt{2} \, n \pi k}{e^{2\pi kR} -1} \,    (A^{\aL}_y - A^{\aR}_y)   
\label{largeGT2}
\eneq
for $a=1,2,3$ so that the Neumann boundary condition for $A^{\hat a}_y$
$(a=1,2,3)$ is preserved only if $A^{\aL}_y =A^{\aR}_y$ at the boundary.
Since $A^{\aL}_y$ obeys the Dirichlet boundary condition,  $A^{\aR}_y$
must obey the Dirichlet boundary condition as well.

With the boundary condition in Table~\ref{bd_gauge}, 
 the gauge symmetry  $SO(5)\times\ubl$ in the bulk 
 is reduced to $SO(4)\times\ubl$ at the TeV brane and 
to $\suL\times\uy$ at the Planck brane.
The resultant symmetry of the theory is $\suL \times U(1)_Y$,
which is  subsequently broken to $U(1)_{\rm EM}$ by nonvanishing  
$A^{\hat{a}}_y$ ($\hat{a}=1,2,3,4$)  or $\thH$.
The weak hypercharge $Y$ is given by  $Y=T^{\nR}+q_{\rm B-L}/2$. 

The boundary conditions of $A^{1_{\rm R}}_\mu$, 
$A^{2_{\rm R}}_\mu$, and   $A^{\prime 3_{\rm R}}_\mu$ 
are changed from N to D on the Planck brane, if  additional  dynamics 
on the Planck brane spontaneously breaks $\suR\times\ubl$  to $\uy$ 
at relatively high energy  scale~$M$,  say,  near the Planck scale~$M_{\rm Pl}$
so that $A^{1_{\rm R}}_\mu$, 
$A^{2_{\rm R}}_\mu$, and   $A^{\prime 3_{\rm R}}_\mu$ have masses of 
$O(M)$ on the Planck brane.  
Below the TeV scale, 
the mass terms on the Planck brane strongly suppress the boundary values of
these fields, effectively changing  the boundary conditions from N to D 
at the Planck brane.   With the underlying gauge invariance
it is expected that the tree unitarity for the gauge boson 
scatterings~\cite{Csaki3,Sakai,Chivukula} is 
preserved with these boundary conditions.

\ignore{
We note that when masses are induced by spontaneous symmetry breaking on the
Planck brane, well-controled ultra-violet behavior of the gauge bosons is not 
spoiled so that the finiteness of the 4D Higgs boson mass at the one loop level, for instance,
is expected to be maintained.  It has been shown recently that the requirement of the 
tree level unitarity constrains boundary conditions satisfied by gauge 
bosons.\cite{Sakai,Higgsless,Chivukula} \ 
With the underlying mechanism of spontaneous symmetry breaking,  the effective
boundary conditions in Table I are expected to preserve the tree level 
unitarity.
}

\begin{table}[t]
\caption{Boundary conditions for the gauge fields. 
$\aL,\aR=1,2,3$ and $\hat{a}=1,2,3,4$. The notation~(D,N), for example, 
denotes the Dirichlet boundary condition at $y=0$ and the Neumann boundary 
condition at $y=\pi R$. }  
\label{bd_gauge}
\begin{center}
\begin{tabular}{ccccc|ccccc} \hline\hline 
 \rule[-2mm]{0mm}{7mm} $A^{\aL}_\mu$ & $A^{1,2_{\rm R}}_\mu$ & 
 $A^{\prime \nR}_\mu$ & $A^Y_\mu$ & $A^{\hat{a}}_\mu$ &
 $A^{\aL}_y$ & $A^{1,2_{\rm R}}_y$ & 
 $A^{\prime \nR}_y$ & $A^Y_y$ & $A^{\hat{a}}_y$ \\ \hline 
 (N,N) & (D,N) & (D,N) & (N,N) & (D,D)  &
 (D,D) & (D,D) & (D,D) & (D,D) & (N,N) \\ \hline
 \end{tabular}
\end{center}
\end{table}

\section{Spectrum and mode functions of gauge bosons}

In this section we derive the spectrum and  mode functions 
of gauge fields with the boundary conditions listed in Table~\ref{bd_gauge}. 
Although such quantities have been well discussed in many papers 
in the case of vanishing $\thH$ (see Refs.~\citen{GP,Chang,Flachi} for example), 
the case of nonzero $\thH$ becomes highly nontrivial 
as the boundary conditions of 5D fields are twisted by the angle~$\thH$, 
\ie, they are no longer either the ordinary Neumann or Dirichlet boundary conditions. 
In fact the $SU(3)$ model has been analysed in Ref.~\citen{HNSS} 
where it is found that the wave functions have nontrivial $\thH$ dependence. 
Here and in the next section we provide systematic KK analysis and obtained 
the full spectrum and the wave functions 
in  the $SO(5)\times\ubl$ model 
for a general value of $\thH$. 
The results obtained here are used to estimate variouos coupling constants 
in Sec.~\ref{couplings}. 

\subsection{General solutions in the bulk}

The basic procedure is the same as in our previous papers.\cite{SH1,HNSS} \  
We employ the background field method, separating ~$A_M$ ($B_M$) 
into the classical part~$A^{\rm c}_M$ ($B^{\rm c}_M$) 
and the quantum part~$A^{\rm q}_M$ ($B^{\rm q}_M$);
$A_M = A^{\rm c}_M+A^{\rm q}_M$ and
$ B_M = B^{\rm c}_M+B^{\rm q}_M$.
It is convenient go over to 
the conformal coordinate~$z\equiv e^{\sgm(y)}$ for the fifth dimension;
\bea
 ds^2 \eql \frac{1}{z^2}\brc{\eta_{\mu\nu}dx^\mu dx^\nu+\frac{dz^2}{k^2}}, 
 \nonumber\\
 \noalign{\kern 5pt}
 \der_y \eql kz\der_z, \;\;\;
 A_y = kzA_z, \;\;\; B_y = kzB_z. 
\eea
The boundaries are located at $z=1$ and  $\zp\equiv e^{k\pi R}$. 
The gauge-fixing functions are chosen \cite{Oda1} as 
\bea
 f^{(A)}_{\rm gf} \eql 
 z^2 \bigg\{ \eta^{\mu\nu}\cD^{\rm c}_\mu A^{\rm q}_\nu
 +\xi k^2 z  \cD^{\rm c}_z \Big( \frac{1}{z} A^{\rm q}_z \Big) \bigg\} ~~, \cr
 \noalign{\kern 10pt}
 f^{(B)}_{\rm gf} \eql 
 z^2 \bigg\{ \eta^{\mu\nu}\der_\mu B^{\rm q}_\nu
 +\xi k^2 z  \der_z  \Big( \frac{1}{z} B^{\rm q}_z \Big) \bigg\}
 \label{f_gf}
\eea
where $\cD^{\rm c}_M A^{\rm q}_N \equiv 
\der_M A^{\rm q}_N-ig_A[A^{\rm c}_M,A^{\rm q}_N]$. 

The quadratic terms for the gauge and ghost fields are simplified for $\xi=1$, 
\bea
 S \eql \int d^4x \,  \frac{dz}{kz} \, \bigg[
 \tr\brc{\eta^{\mu\nu}A^{\rm q}_\mu \brkt{\Box+k^2 {\cP}_4} A^{\rm q}_\nu
 +k^2 A^{\rm q}_z \brkt{\Box+k^2 {\cP}_z } A^{\rm q}_z
 +\cL_{\rm gh}^{(A)}}     \cr
\noalign{\kern 10pt}
 &&\hskip 2cm 
 \left.+\eta^{\mu\nu}B^{\rm q}_\mu\brkt{\Box+k^2 {\cP}_4}B^{\rm q}_\nu
 +k^2B^{\rm q}_z\brkt{\Box+k^2 {\cP}_z}B^{\rm q}_z
 +\cL_{\rm gh}^{(B)}\right], 
 \label{action3}
\eea
where 
\be
\Box\equiv\eta^{\mu\nu}\der_\mu\der_\nu ~~,~~
  {\cP}_4 \equiv z\cD^{\rm c}_z\frac{1}{z}\cD^{\rm c}_z ~~,~~
  {\cP}_z \equiv \cD^{\rm c}_z z\cD^{\rm c}_z\frac{1}{z} ~~. 
\ee
Here we have taken $A^{\rm c}_\mu=0$, respecting the 4D Poincar\'{e} 
symmetry.  The surface terms at the boundaries at $z=1, z_\pi$ vanish 
thanks to the boundary conditions for each field. 

The linearized equations of motion for $A_M$ become 
\bea
 \Box A^{\rm q}_\mu+k^2z\cD^{\rm c}_z\frac{1}{z}\cD^{\rm c}_z A^{\rm q}_\mu
 \eql 0, \nonumber\\
 \Box A^{\rm q}_z+k^2\cD^{\rm c}_z z\cD^{\rm c}_z\frac{1}{z}A^{\rm q}_z 
 \eql 0, 
 \label{lin_EOM1}
\eea
and those for $B_M$ have similar forms. 
The classical background is taken to be 
$A^{\rm c}_z = v z T^{\hat{4}}$ ($v$: real constant) below. 

We move to a twisted basis by a gauge transformation;\footnote{
We define $\Omg(z)$ so that $\Omg(\zp)=1_4$ 
in contrast to our previous work~\cite{SH1,HNSS} where it is defined as 
$\Omg(1)=1_4$. }
\bea
 \tl{A}_M \defa \Omg A^{\rm q}_M \Omg^{-1}, 
 \;\;\;
 \tl{B}_M \equiv B^{\rm q}_M, \nonumber\\
\noalign{\kern 5pt}
 \Omg(z) \defa \exp\brc{ig_A\int_z^{\zp}dz^\prime\;A^{\rm c}_z(z^\prime)}. 
 \label{def_Omg}
\eea
As shown in Refs.~\citen{YH2,HHHK,YHscgt1},  sets of boundary conditions related by gauge 
transformations form equivalence classes and all sets in each equivalence class
are physically equivalent.  Dynamics of the Yang-Mills AB phase $\thH$ 
guarantees the equivalence.
In the twisted basis the classical background of the gauge fields vanishes 
so that the linearized equations of motion reduce to 
free field equations, 
while the boundary conditions become more involved. 
\bea
 \Box \tl{A}_\mu+k^2\brkt{\der_z^2-\frac{1}{z}\der_z}\tl{A}_\mu \eql 0 ~~,~~
 \Box \tl{A}_z+k^2\brkt{\der^2-\frac{1}{z}\der_z+\frac{1}{z^2}}\tl{A}_z 
 = 0 ~~, \nonumber\\
 \noalign{\kern 5pt}
 \Box \tl{B}_\mu+k^2\brkt{\der_z^2-\frac{1}{z}\der_z}\tl{B}_\mu \eql 0 ~~,~~
 \Box \tl{B}_z+k^2\brkt{\der^2-\frac{1}{z}\der_z+\frac{1}{z^2}}\tl{B}_z 
 = 0 ~~. 
\eea
Thus the equations for eigenmodes with a mass eigenvalue~$m_n=k\lmd_n$ are 
\bea
 \brc{\frac{d^2}{dz^2}-\frac{1}{z}\frac{d}{dz}+\lmd_n^2}\tl{h}^I_{A,n}
 \eql = 0 ~~, \nonumber\\
 \brc{\frac{d^2}{dz^2}-\frac{1}{z}\frac{d}{dz}+\frac{1}{z^2}+\lmd_n^2}
 \tl{h}^I_{\vph,n}
 \eql = 0 ~~. 
 \label{md_eq1}
\eea
With these eigenfunctions the gauge potentials are expanded as 
\bea
 \tl{A}^I_\mu(x,z) \eql \sum_n \tl{h}^I_{A,n}(z)A^{(n)}_\mu(x) ~~,~~
 \tl{A}^I_z(x,z) = \sum_n \tl{h}^I_{\vph,n}(z)\vph^{(n)}(x) ~~, \cr
 \tl{B}_\mu(x,z) \eql \sum_n \tl{h}^B_{A,n}(z)A^{(n)}_\mu(x) ~~,~~
 \tl{B}_z(x,z) = \sum_n \tl{h}^B_{\vph,n}(z)\vph^{(n)}(x) ~~. 
 \label{gauge_md_ex}
\eea
The general solutions to Eq.(\ref{md_eq1}) are expressed in terms of 
the Bessel functions as 
\bea
 \tl{h}^{\bar{I}}_{A,n}(z) \eql 
 z\brc{\alp^{\bar{I}}_{A,n}J_1(\lmd_n z)+\bt^{\bar{I}}_{A,n}Y_1(\lmd_n z)} ~~, 
 \nonumber\\
 \tl{h}^{\bar{I}}_{\vph,n}(z) \eql 
 z\brc{\alp^{\bar{I}}_{\vph,n}J_0(\lmd_n z)
 +\bt^{\bar{I}}_{\vph,n}Y_0(\lmd_n z)}  ~~,
  \label{gen_sol1}
\eea
where $\bar{I}=(\aL,\aR,\hat{a},B)$ and 
$\alp_n$'s and $\bt_n$'s are constants to be determined 
by the boundary conditions. 

\subsection{Mass eigenvalues and mode functions}

To determine the eigenvalues~$\lmd_n$'s and 
the corresponding mode functions~(\ref{gen_sol1}), 
we need to take into account the boundary conditions 
listed in Table~\ref{bd_gauge}. 
In this subsection we mainly examine  the 4D components of 
the gauge fields~($A_\mu$, $B_\mu$). 
The mass spectrum and the mode functions for the extra-dimensional components 
$(A_z, B_z)$ are examined in the next subsection. 

At the boundaries
\beqn
&&\hskip -1cm 
 \der_z A^{\aL}_\mu = \der_z A^Y_\mu =0 ~~,~~
  A^{1,2_{\rm R}}_\mu = A^{\prime \nR}_\mu = A^{\hat{a}}_\mu = 0  
  \hskip 1cm  \hbox{at }z=1  ~, 
\label{UVbd_cond}  \\
\noalign{\kern 5pt}
&&\hskip -1cm 
\der_z A^{\aL}_\mu = \der_z A^{\aR}_\mu = \der_z B_\mu = 0 ~~,~~
A^{\hat{a}}_\mu = 0 
  \hskip 1cm  \hbox{at }z=z_\pi ~. 
\label{IRbd_cond}
\eeqn
We translate these conditions into those 
in the twisted basis~($\tl{A}_M$, $\tl{B}_M$). 
Among the extra-dimensional components of the gauge fields, only $A^{\hat{a}}_z$ 
can have non-vanishing vacuum expectation values (VEV).
With the residual $\suL\times\uy$ symmetry, 
we can restrict ourselves to 
\be
 A^{\rm c}_z = vzT^{\hat{4}}, \label{Az_vev}
\ee
where a constant $v$ is  related to $\thH$ by 
\be
 \thH = \frac{g_A v}{2\sqrt{2}}(\zp^2-1) ~. 
\ee
The potential has a classical flat direction along $\thH$. 
The value for $\thH$ is determined at the quantum level. 
Using (\ref{Az_vev}), the gauge transformation matrix~$\Omg$ 
defined in (\ref{def_Omg}) is calculated as 
\bea
 \Omg(z) = \exp\brc{ i\frac{1}{2}\tht(z)  \big(2\sqrt{2} \, T^{\hat{4}} \big) } 
 = \mtrx{\cth 1_2}{i\sth 1_2}{i\sth 1_2}{\cth 1_2}, 
 \label{expr_Omg}
\eea
where 
\be
 \tht(z) \equiv \frac{g_A}{\sqrt{2}}  \int_z^{\zp}dz^\prime \; 
 A^{{\rm c}\hat{4}}_z(z^\prime) 
 =\frac{g_A v}{2\sqrt{2}}(\zp^2-z^2) = \thH\frac{\zp^2-z^2}{\zp^2-1} ~, 
 \label{def_tht}
\ee
and $\cth\equiv\cos\frac{1}{2}\tht(z)$, $\sth\equiv\sin\frac{1}{2}\tht(z)$. 
Thus the relation between $A_M$ and $\tl{A}_M$ in Eq.(\ref{def_Omg}) 
can be written as 
\bea
 \brkt{\begin{array}{c} \tl{A}^{\aL}_M \\ \tl{A}^{\aR}_M \\ 
 \tl{A}^{\hat{a}}_M  \end{array}}
 \eql \brkt{\begin{array}{ccc} \cth^2 & \sth^2 & \sqrt{2}\sth\cth \\
 \sth^2 & \cth^2 & -\sqrt{2}\sth\cth \\ -\sqrt{2}\sth\cth & \sqrt{2}\sth\cth & 
 \cth^2-\sth^2 \end{array}}
 \brkt{\begin{array}{c} A^{{\rm q}\aL}_M \\ A^{{\rm q}\aR}_M \\ 
 A^{{\rm q}\hat{a}}_M  \end{array}}, \nonumber\\
 \tl{A}^{\hat{4}}_M \eql A^{{\rm q}\hat{4}}_M ~, 
 \label{rel_to_tl}
\eea
where $a=1,2,3$. 

The boundary conditions~(\ref{UVbd_cond})  and (\ref{IRbd_cond}) can be 
rewritten in terms of $\tl{A}_\mu$ by using this relation. 
The condition~(\ref{IRbd_cond}) determines 
the ratios between $\alp_{A,n}$'s and $\bt_{A,n}$'s in Eq.(\ref{gen_sol1}) 
so that the mode functions have the following forms. 
\bea
 \tl{h}^{\aL}_{A,n}(z) \eql C^{\aL}_{A,n} z F_{1,0}(\lmd_n z,\lmd_n\zp) ~, 
 \nonumber\\
 \tl{h}^{\aR}_{A,n}(z) \eql C^{\aR}_{A,n} z F_{1,0}(\lmd_n z,\lmd_n\zp) ~, 
 \nonumber\\
 \tl{h}^{\hat{a}}_{A,n}(z) \eql C^{\hat{a}}_{A,n} 
 z F_{1,1}(\lmd_n z,\lmd_n\zp) ~, \nonumber\\
 \tl{h}^B_{A,n}(z) \eql C^B_{A,n} z F_{1,0}(\lmd_n z,\lmd_n\zp) ~.
 \label{sol2}
\eea
Here the functions~$F_{\alp,\bt}(u,v)$ are defined in (\ref{def_Fs}). 
The mass eigenvalue~$\lmd_n$ and the coefficients~$C_n$'s 
are determined by the remaining boundary 
condition~(\ref{UVbd_cond}), which amount to
\be
 \brkt{\begin{array}{ccc} \cw^2\lmd_nF_{0,0} & \sw^2\lmd_nF_{0,0} &
 -\sqrt{2}\sw\cw F_{0,1} 
 \\ \sw^2\lmd_nF_{1,0} & \cw^2\lmd_nF_{1,0} & \sqrt{2}\sw\cw F_{1,1} \\
 \sqrt{2}\sw\cw\lmd_nF_{1,0} & -\sqrt{2}\sw\cw\lmd_nF_{1,0} & (\cw^2-\sw^2)F_{1,1} 
 \end{array}}\brkt{\begin{array}{c} C^{\aL}_{A,n} \\ C^{\aR}_{A,n} 
 \\ C^{\hat{a}}_{A,n}\end{array}} = 0 ~, 
 \label{charged_bdcd}
\ee
for $a=1,2$, and 
\be
 \brkt{\begin{array}{cccc} \cw^2\lmd_nF_{0,0} & \sw^2\lmd_nF_{0,0} &
 -\sqrt{2}\sw\cw F_{0,1} & 0 \\ 
 \cph\sw^2\lmd_nF_{1,0} & \cph\cw^2\lmd_nF_{1,0} & \sqrt{2}\cph\sw\cw F_{1,1} 
 & -\sph\lmd_nF_{1,0} \\ 
 \sqrt{2}\sw\cw\lmd_nF_{1,0} & -\sqrt{2}\sw\cw\lmd_nF_{1,0} 
 & (\cw^2-\sw^2)F_{1,1} & 0 \\ 
 \sph\sw^2\lmd_nF_{0,0} & \sph\cw^2\lmd_nF_{0,0} & 
 \sqrt{2}\sph\sw\cw F_{0,1} & \cph\lmd_nF_{0,0} \end{array}}
 \brkt{\begin{array}{c} C^{\nL}_{A,n} \\ C^{\nR}_{A,n} \\ C^{\hat{3}}_{A,n} \\
 C^B_{A,n} \end{array}} = 0 ~, 
 \label{neutral_bdcd}
\ee
\be
 F_{1,1}C^{\hat{4}}_{A,n} = 0 ~, 
 \label{singlet_bdcd}
\ee
where $\sw\equiv\sin\frac{1}{2}\thH$, $\cw\equiv\cos\frac{1}{2}\thH$. 
Here and henceforth, $F_{\alp,\bt}$ without the argument denotes 
$F_{\alp,\bt}(\lmd_n,\lmd_n\zp)$. 

$U(1)$ subgroup remains unbroken 
for any value of nonzero $\thH$, which is identified with the electromagnetic 
gauge group~$\uem$. 
The gauge fields are classified in three sectors,  the charged sector 
\be
 (A^{\chL}_M,A^{\chR}_M, A_M^{\hat{\pm}}) \equiv \frac{1}{\sqrt{2}}
 (A^{1_{\rm L}}_M\pm iA^{2_{\rm L}}_M, 
 A^{1_{\rm R}}_M\pm iA^{2_{\rm R}}_M,A^{\hat{1}}_M\pm iA^{\hat{2}}_M) ~, 
 \label{def_charged}
\ee
the neutral sector
\be
 (A^{\nL}_M,A^{\nR}_M,A^{\hat{3}}_M,B_M) ~, \label{def_neutral}
\ee
and the `singlet' sector~$A^{\hat{4}}_M$. 
The latter two sectors are neutral under $\uem$. 
The orthonormal relations among the mode functions are 
\bea
 \int_1^{\zp}\frac{dz}{kz}\brc{\tl{h}^{\chL}_{A,n}\tl{h}^{\chL}_{A,l}
 +\tl{h}^{\chR}_{A,n}\tl{h}^{\chR}_{A,l}+\tl{h}^{\hat{\pm}}_{A,n}
 \tl{h}^{\hat{\pm}}_{A,l}} \eql \dlt_{n,l} ~, \nonumber\\
 \int_1^{\zp}\frac{dz}{kz}\brc{\tl{h}^{\nL}_{A,n}\tl{h}^{\nL}_{A,l}
 +\tl{h}^{\nR}_{A,n}\tl{h}^{\nR}_{A,l}+\tl{h}^{\hat{3}}_{A,n}
 \tl{h}^{\hat{3}}_{A,l}+\tl{h}^B_{A,n}\tl{h}^B_{A,l}} \eql \dlt_{n,l} ~, 
 \nonumber\\
 \int_1^{\zp}\frac{dz}{kz}\;\tl{h}^{\hat{4}}_{A,n}\tl{h}^{\hat{4}}_{A,l} 
 \eql \dlt_{n,l} ~. 
 \label{orthonormal_gauge}
\eea

\subsubsection{Charged sector 
$ (A^{\chL}_\mu, A^{\chR}_\mu,  A_\mu^{\hat{\pm}})$} \label{charged_sector}

In order for  nontrivial solutions to Eq.\ (\ref{charged_bdcd}) 
to exist, the determinant of the $3\times 3$ 
matrix must vanish, which leads to 
\be
 F_{1,0}\brc{\pi^2 \lmd_n^2\zp F_{0,0}F_{1,1}
 - 2 \sin^2\thH} = 0 ~ . 
\ee
Once the spectrum $\lmd_n$ is determined by the above equation, 
the corresponding $C_{A,n}$'s are fixed by 
(\ref{charged_bdcd}) with the normalization 
condition~(\ref{orthonormal_gauge}). 

There are two cases for the mass spectrum. 

\noindent
{\bf \under{Case 1}:}  $F_{1,0} = 0$

There is no massless mode and 
the lightest mode has a mass of $\cO(\mKK)$, 
where the Kaluza-Klein (KK) mass scale $\mKK$ is given by 
\be
 \mKK \equiv \frac{k\pi}{\zp-1} ~.
\ee
The coefficients~$C^I_{A,n}$ ($I=\chL,\chR,\hat{\pm}$) 
in the mode functions are given by 
\bea
 C^{\chL}_{A,n} \eql  (1-\cos\thH) \hat C_1 ~~,  \cr
\noalign{\kern 5pt}
 C^{\chR}_{A,n} \eql - (1+\cos\thH) \hat C_1 ~~, \cr
\noalign{\kern 5pt}
 C^{\hat{\pm}}_{A,n} \eql 0 ~~, \cr
\noalign{\kern 5pt}
 \hat C_1 \eql   \frac{\sqrt{k}} {\sqrt{1+\cos^2\thH}}
 \brc{\frac{4}{\pi^2\lmd_n^2}-F_{0,0}^2}^{-1/2} ~~.
 \label{charged_Cs1}
\eea
The mass spectrum is independent of $\thH$ and 
is the same as  for the modes with the boundary condition~(D,N) 
at $\thH=0$. 
For nonzero $\thH$, however, 
the above modes do not have definite $Z_2$-parities 
since components with different boundary conditions  mix with each other. 
This can be seen explicitly from the fact that 
the mode functions have nontrivial $\thH$-dependences. 

\noindent
{\bf \under{Case 2}:}  $\pi^2\lmd_n^2\zp F_{0,0}F_{1,1}=2\sin^2\thH$

In this case  the mass spectrum depends on $\thH$. 
The lightest mode is massless at $\thH=0$, 
while it acquires a nonvanishing mass when $\thH\neq 0$. 
The lightest mode is identified with the $W$ boson. 
The coefficients in the mode functions are given by
\beqn
&&\hskip -1cm 
C^{\chL}_{A,n} =  (1+\cos\thH) ~ \hat C_2 ~, \cr
\noalign{\kern 5pt}
&&\hskip -1cm 
C^{\chR}_{A,n} = (1- \cos\thH) ~ \hat C_2  ~, \cr
\noalign{\kern 5pt}
&&\hskip -1cm 
C^{\hat{\pm}}_{A,n} =  - \sqrt{2}\sin\thH 
   \frac{F_{1,0}}{F_{1,1}}  ~ \hat C_2   ~ , \cr
\noalign{\kern 5pt}
&&\hskip -1cm 
\hat C_2 = \frac{\sqrt{k}}{\sqrt{1+\cos^2\thH}}
 \brc{\frac{4}{\pi^2\lmd_n^2}+\frac{\pi^2\lmd_n^2\zp^2F_{1,0}^2F_{0,0}^2}
 {\sin^2\thH(1+\cos^2\thH)}-\frac{2F_{1,0}^2}{1+\cos^2\thH}
 -\frac{2F_{0,0}^2}{\sin^2\thH}}^{-1/2} .
 \label{charged_Cs2}
\eea
One comment is in order about the behavior of  $\hat C_2$ in the 
$\thH \go 0$ limit.  For the KK excited states, or modes
with $\lim_{\thH \go 0} \lambda_n \not= 0$, either $F_{0,0}$ or
$F_{1,1}$ is $\cO (\thH^2)$.    $C^{\chL}_{A,n}$
becomes dominant for  the modes with $F_{0,0}=\cO (\thH^2)$,
while $C^{\hat{\pm}}_{A,n}$ becomes dominant 
for  the modes with $F_{1,1}=\cO (\thH^2)$.
For the zero mode $\lambda_0 = \cO (\thH)$ and 
$\hat C_2 \simeq \sqrt{\pi} \lambda_0 / 4 \sqrt{R}$.

\subsubsection{Neutral sector 
$ (A^{\nL}_\mu,A^{\nR}_\mu,A^{\hat{3}}_\mu,B_\mu)$} \label{neutral_sector}

The determinant of the $4\times 4$ matrix in (\ref{neutral_bdcd})    must vanish, 
which leads to  
\be
 \lmd_n F_{0,0}F_{1,0} \brc{ \pi^2 \lmd_n^2\zp F_{0,0}F_{1,1}
 - 2  ( 1+\sph^2) \sin^2 \thH }=0 ~~.
 \label{neutral-spec1}
\ee
$\sph$ is defined in (\ref{def_vph}).  
This determines the mass spectrum. 
Once $\lmd_n$ is determined, 
the coefficients~$C^{\bar{I}}_{A,n}$ ($\bar{I}=\nL,\nR,\hat{3},B$) 
in mode functions are 
fixed by (\ref{neutral_bdcd}) with the normalization 
condition~(\ref{orthonormal_gauge}). 

The neutral sector is classified into  three cases. 

\noindent
{\bf \under{Case 1}:}  $\lmd_n F_{0,0} = 0$  

The massless mode ($\lmd_0=0$) identified with the photon for the 
unbroken $\uem$ has a constant mode function
\bea
 \tl{h}^{\nL}_{A,0} \eql \tl{h}^{\nR}_{A,0} 
 = \frac{\sph}{\sqrt{(1+\sph^2)\pi R}} ~~, 
 \nonumber\\
 \tl{h}^{\hat{3}}_{A,0} \eql 0 ~~, \;\;\;\;\;
 \tl{h}^B_{A,0} = \frac{\cph}{\sqrt{(1+\sph^2)\pi R}} ~~. 
 \label{zero_photon}
\eea
The  massive modes have
\bea
 \tl{h}^{\nL}_{A,n}(z) \eql \tl{h}^{\nR}_{A,0}(z) = 
 \frac{\sqrt{2k}\sph}{\sqrt{1+\sph^2}}
 \brc{\frac{4}{\pi^2\lmd_n^2}-F_{1,0}^2}^{-1/2} 
 F_{1,0}(\lmd_n z,\lmd_n\zp) ~~, \nonumber\\
 \tl{h}^{\hat{3}}_{A,n}(z) \eql 0 ~~, \;\;\;\;\;
 \tl{h}^B_{A,n}(z) = \frac{\cph}{\sph}\, \tl{h}^{\nL}_{A,n}(z)~~. 
 \label{nonzero_photon}
\eea
(\ref{zero_photon}) can be obtained also from (\ref{nonzero_photon})
by taking the limit of $\lmd_n\to 0$. 
Note that the mass spectrum and the mode functions in the photon sector 
are independent of $\thH$. 
In fact we can extract the photon sector from the neutral sector 
by the following field redefinition. 
\be
 \brkt{\begin{array}{c} A^{3_V}_\mu \\ A^{3_A}_\mu \\ A^\gm_\mu 
 \end{array}} = \frac{1}{\sqrt{2(1+\sph^2)}}
 \brkt{\begin{array}{ccc} \cph & \cph & -2\sph \\ 
 \sqrt{1+\sph^2} & -\sqrt{1+\sph^2} & 0 \\
 \sqrt{2}\sph & \sqrt{2}\sph & \sqrt{2}\cph \end{array}}
 \brkt{\begin{array}{c} A^{\nL}_\mu \\ A^{\nR}_\mu \\ B_\mu 
 \end{array}}. 
\ee
The photon field~$A^\gm_\mu$ does not mix with 
the other components~$(A^{3_V}_\mu,A^{3_A}_\mu)$ 
under the $\Omega$-rotation in (\ref{expr_Omg}).

\noindent
{\bf \under{Case 2}:}  $F_{1,0} = 0$  

The equation that determines 
the mass spectrum is the same as in the case~1 in the charged sector. 
The coefficients in the mode functions are given by
\bea
 C^{\nL}_{A,n} \eql  (1-\cos\thH) ~ \hat C_3 ~~, \cr
\noalign{\kern 5pt}
 C^{\nR}_{A,n} \eql  -(1 + \cos\thH)  ~ \hat C_3 ~~, \cr
\noalign{\kern 5pt}
 C^{\hat{3}}_{A,n} \eql 0 ~~, \cr
\noalign{\kern 5pt}
 C^B_{A,n} \eql  2\tph\cos\thH  ~ \hat C_3 ~~,  \cr
 \noalign{\kern 5pt}
 \hat C_3 \eql   \frac{\sqrt{2k}}
 {\sqrt{1+(1+4\tph^2)\cos^2\thH}}
 \brc{\frac{4}{\pi^2\lmd_n^2}-F_{0,0}^2}^{-1/2} ~~, 
 \label{neutral_Cs2}
\eea
where $\tph \equiv \sph/\cph$.

\noindent
{\bf \under{Case 3}:}
  $\pi^2\lmd_n^2\zp F_{0,0}F_{1,1} = 2(1+\sph^2)\sin^2\thH$ 

In this case the mass spectrum depends on $\thH$. 
The lightest mode becomes massless at  $\thH =0$ while it acquires 
a nonzero mass when $\thH\neq 0$. 
The mode is identified with the $Z$ boson. 
The coefficients in the mode functions are given by 
\bea
 C^{\nL}_{A,n} \eql   \big\{ \cph^2+\cos\thH(1+\sph^2) \big\} ~\hat C_4 ~~, \cr
 \noalign{\kern 5pt}
 C^{\nR}_{A,n} \eql  \big\{ \cph^2- \cos\thH(1+\sph^2) \big\} ~\hat C_4 ~~, \cr
 \noalign{\kern 5pt} 
  C^{\hat{3}}_{A,n} \eql   - \sqrt{2} (1+\sph^2)\sin\thH  \, 
      \frac{F_{1,0}}{F_{1,1}}   ~\hat C_4 ~~, \cr
 \noalign{\kern 5pt}
 C^B_{A,n} \eql -\frac{2\sph\cph}{\cph^2+\cos\thH(1+\sph^2)}C^{\nL}_{A,n}.  \cr
 \noalign{\kern 5pt}
 \hat C_4 \eql   \frac{\sqrt{k} } {\sqrt{2(1+\sph^2)}\cos\chi}
 \brc{\frac{4}{\pi^2\lmd_n^2}
 +\frac{\pi^2\lmd_n^2\zp^2F_{1,0}^2F_{0,0}^2}{\sin^22\chi}
 -\frac{F_{1,0}^2}{\cos^2\chi}-\frac{F_{0,0}^2}{\sin^2\chi}}^{-1/2}  , \cr
\noalign{\kern 5pt}
&&
\sin^2 \chi \equiv \frac{1+\sph^2}{2} \sin^2 \thH ~~.  
\label{neutral_Cs3}
\eea

Note that the $4\times 4$ matrix in (\ref{neutral_bdcd}) reduces to 
a direct sum of $3\times 3$ and $1\times 1$ matrices 
and the former is identical to the $3\times 3$ matrix 
in (\ref{charged_bdcd}) if we set $(\sph,\cph)=(0,1)$. 
Thus the spectrum and the mode functions 
of the charged sector~~$(A^{\chL}_\mu,A^{\chR}_\mu,A^{\hat{3}}_\mu)$ 
are reproduced from those of $(A^{\nL}_\mu,A^{\nR}_\mu,A^{\hat{3}}_\mu)$ 
by setting $(\sph,\cph)=(0,1)$.

\subsubsection{Singlet sector $A^{\hat{4}}_\mu$}
Finally there is the singlet sector $A^{\hat{4}}_\mu$. 
There is no zero mode in this sector.
From the normalization condition~(\ref{orthonormal_gauge}), 
the coefficient is determined  as 
\be
 C^{\hat{4}}_{A,n} = \sqrt{2k}\brc{\frac{4}{\pi^2\lmd_n^2}-F_{0,1}^2}^{-1/2}. 
\ee

In the gauge sector, there are some classes of K.K. modes whose spectra 
are independent of $\thH$, \ie, the case 1 in the charged 
sector, the cases 1 and 2 in the neutral sector, 
and the singlet sector. 
In all these cases, the mode functions do not have nonzero components for 
$T^{\hat{a}}$ ($a=1,2,3$, or $\pm,3$). 
This means that the modes in these classes do not have 
nonvanishing couplings to the Higgs field~$H=\vph^{(0)}$, 
which reflects the $\thH$-independence of the mass spectra. 
(The corresponding coupling constants are expressed like (\ref{def_WWH}) 
in Sect.~\ref{WWHcp}.)
\ignore{
The $\thH$-independence of the spectrum is also understood 
from the fact that they do not have 
nonvanishing couplings to the Higgs field~$\vph^{(0)}$. 
}

\subsection{Spectrum and mode functions of gauge scalars}

In this subsection the spectrum and mode functions for the extra-dimensional 
components of gauge potentials, or gauge scalars, are examined. 
The boundary conditions for $A_z$ and $B_z$ are given by
\be
 A_z^{\aL} = A_z^{\aR} = B_z =0 ~~,~~
 \der_z\brkt{\frac{A_z^{\hat{a}}}{z}}=0  
 \label{bd_cd:Az} 
\ee
at both boundaries. 
The conditions at $z=\zp$ determine the ratios between $\alp_{\vphi, n}$'s 
and $\bt_{\vphi, n}$'s 
in Eq.(\ref{gen_sol1}) so that the mode functions have the following forms. 
\bea
 \tl{h}_{\vph,n}^{\aL}(z) \eql C^{\aL}_{\vph,n}z F_{0,0}(\lmd_n z,\lmd_n\zp) ~, 
 \nonumber\\
 \tl{h}_{\vph,n}^{\aR}(z) \eql C^{\aR}_{\vph,n}z F_{0,0}(\lmd_n z,\lmd_n\zp) ~, 
 \nonumber\\
 \tl{h}_{\vph,n}^{\hat{a}}(z) \eql C^{\hat{a}}_{\vph,n}z F_{0,1}(\lmd_n z,\lmd_n\zp) ~, 
 \nonumber\\
 \tl{h}^B_{\vph,n}(z) \eql C^B_{\vph,n}z F_{0,0}(\lmd_n z,\lmd_n\zp) ~. 
 \label{md_fct:Az}
\eea

To treat gauge scalars, it is convenient to  define 
\be
 A^{\aLR{\pm}}_z \equiv \frac{1}{\sqrt{2}}\brkt{
 A^{\aL}_z\pm A^{\aR}_z} ~~,
\label{def_LR}
\ee
in terms of which  Eq(\ref{rel_to_tl}) can be rewritten  as 
\bea
 \tl{A}^{\aLR{+}}_z \eql A^{{\rm q}\aLR{+}}_z ~, \nonumber\\
\noalign{\kern 10pt}
 \vct{\tl{A}^{\aLR{-}}_z}{\tl{A}^{\hat{a}}_z} \eql 
 \mtrx{\cos\tht}{\sin\tht}{-\sin\tht}{\cos\tht}
 \vct{A^{{\rm q}\aLR{-}}_z}{A^{{\rm q}\hat{a}}_z} ~, \nonumber\\
\noalign{\kern 10pt}
 \tl{A}^{\hat{4}}_z \eql A^{{\rm q}\hat{4}}_z ~, 
\eea
where $a=1,2,3$. 
In contrast to the 4D components~$A_\mu^I$ and $B_\mu$, 
the boundary conditions in (\ref{bd_cd:Az}) do not mix 
$\tl{A}^{\aLR{+}}_z$, $(\tl{A}^{\aLR{-}}_z,\tl{A}^{\hat{a}}_z)$ and $B_z$. 
By making use of (\ref{md_fct:Az}), 
the boundary conditions at the Planck brane is rewritten as 
\be
 F_{0,0}C^{\aLR{+}}_{\vph,n} = 0 ~~,  \label{C+:Az}
\ee
\be
 \mtrx{\cos\thH F_{0,0}}{-\sin\thH F_{0,1}}{\sin\thH F_{1,0}}{\cos\thH F_{1,1}}
 \vct{C^{\aLR{-}}_{\vph,n}}{C^{\hat{a}}_{\vph,n}} = 0 ~, 
 \label{C-h:Az}
\ee
for $a=1,2,3$, and 
\be
 \lmd_nF_{1,1}C^{\hat{4}}_{\vph,n} = 0 ~, \label{C4:Az}
\ee
\be
 F_{0,0}C^B_{\vph,n} = 0 ~. \label{CB:Az}
\ee
Here $F_{\alp,\bt} =F_{\alp,\bt} (\lmd_n,\lmd_n\zp)$. 
The orthonormal relations are given by 
\bea
 \int_1^{\zp} \frac{k dz}{z} ~\tl{h}^{\aLR{+}}_{\vph,n}\tl{h}^{\aLR{+}}_{\vph,l}
 \eql \dlt_{n,l} ~, \nonumber\\
\noalign{\kern 5pt}
 \int_1^{\zp} \frac{k dz}{z} ~ \brc{\tl{h}^{\aLR{-}}_{\vph,n}
 \tl{h}^{\aLR{-}}_{\vph,l}+\tl{h}^{\hat{a}}_{\vph,n}\tl{h}^{\hat{a}}_{\vph,l}} 
 \eql \dlt_{n,l} ~, \nonumber\\
\noalign{\kern 5pt}
 \int_1^{\zp} \frac{k dz}{z} ~\tl{h}^{\hat{4}}_{\vph,n}\tl{h}^{\hat{4}}_{\vph,l}
 \eql \dlt_{n,l} ~, \nonumber\\
\noalign{\kern 5pt}
 \int_1^{\zp} \frac{kdz}{z} ~\tl{h}^B_{\vph,n}\tl{h}^B_{\vph,l} \eql \dlt_{n,l} ~. 
 \label{orthonormal_rel:Az}
\eea

It follows from the conditions~(\ref{C+:Az}) and (\ref{C-h:Az})
that the spectra for the charged sector~(\ref{def_charged}) and 
the neutral sector~(\ref{def_neutral})  are degenerate. 
The gauge scalar sector is classified into  four cases 
specified by Eqs.(\ref{C+:Az})-(\ref{CB:Az}). 

\bigskip
\noindent
{\bf \under{Case 1}:  Singlet sector I $A^{\aLR{+}}_z$}

For the $A^{\aLR{+}}_z$  components
the mass spectrum is determined by 
\be
 F_{0,0} = 0 ~. 
\ee
There are no zero-modes and the coefficients~$C^{\bar{I}}_{\vph,n}$ 
in the mode functions are determined 
by the normalization condition~(\ref{orthonormal_rel:Az}). 
\bea
 C_{\vph,n}^{\aLR{+}} \eql \sqrt{\frac{2}{k}}\brc{
 \frac{4}{\pi^2\lmd_n^2}-F_{1,0}^2}^{-1/2} ~, \nonumber\\
\noalign{\kern 5pt}
 C_{\vph,n}^{\aLR{-}} \eql C^{\hat{a}}_{\vph,n} = C^B_{\vph,n} = 0 ~, 
\label{scalarC1}
\eea
where $\aLR{\pm}=1,2,3$ and $\hat{a}=1,2,3,4$. 

\bigskip
\noindent
{\bf \under{Case 2}:  Doublet sector $(A^{\aLR{-}}_z,A^{\hat{a}}_z)$}

For $(A^{\aLR{-}}_z,A^{\hat{a}}_z)$  the mass spectrum is determined by 
\be
 F_{0,0}F_{1,1} = \frac{4\sin^2\thH}{\pi^2\lmd_n^2\zp} ~~. 
 \label{scalar_spec2}
\ee
From Eq.(\ref{C-h:Az}) and the normalization condition~(\ref{orthonormal_rel:Az}), 
the mode functions are obtained as 
\bea
 C_{\vph,n}^{\aLR{-}} \eql \sqrt{\frac{2}{k}}\brc{
 \frac{4}{\pi^2\lmd_n^2}+\frac{\pi^2\lmd_n^2\zp^2}{\sin^22\thH}F_{1,0}^2F_{0,0}^2
 -\frac{F_{1,0}^2}{\cos^2\thH}-\frac{F_{0,0}^2}{\sin^2\thH}}^{-1/2}, \nonumber\\
\noalign{\kern 5pt}
 C_{\vph,n}^{\hat{a}} \eql -\tan\thH\frac{F_{1,0}}{F_{1,1}}C^{\aLR{-}}_{\vph,n}, 
 \nonumber\\
\noalign{\kern 5pt}
 C^{\aLR{+}}_{\vph,n} \eql C^B_{\vph,n} = 0. 
 \label{scalarC2}
\eea

\bigskip
\noindent
{\bf \under{Case 3}: Higgs sector $A^{\hat{4}}_z$}

The sector $A^{\hat{4}}_z$ corresponds to the Higgs sector. 
The spectrum is determined by 
\be
 \lmd_n F_{1,1} = 0. 
 \label{scalar_spec3}
\ee
There is a zero-mode, which is identified as the 4D Higgs boson. 
It acquires a nonvanishing finite mass $m_H$ by quantum effects at the one loop
level.  It has been estimated in Refs.~\citen{HM,HNSS} that
$m_H \sim  0.1 \sqrt{\alpha_W} \, k\pi R m_W/  |\sin \thH| $,  
which gives $m_H$ in a range $140 \sim 280$ GeV.
The  mode functions in this case  are given by 
\bea
 \tl{h}^{\hat{4}}_{\vph,0}(z) \eql \sqrt{\frac{2}{k(\zp^2-1)}} ~~z ~~, \cr
 \noalign{\kern 10pt}
 \tl{h}^{\aLR{\pm}}_{\vph,0}(z) \eql \tl{h}^{\hat{a}}_{\vph,0}(z) 
 = \tl{h}^B_{\vph,0}(z) = 0 ~~, 
 \label{scalar_wave3}
\eea
for the zero-mode (the 4D Higgs field), and 
\bea
 C^{\hat{4}}_{\vph,n} \eql \sqrt{\frac{2}{k}}
 \brc{\frac{4}{\pi^2\lmd_n^2}-F_{0,1}^2}^{-1/2} ~~, \nonumber\\
\noalign{\kern 10pt}
 C^{\aLR{\pm}}_{\vph,n} \eql C^{\hat{a}}_{\vph,n} = C^B_{\vph,n} = 0 ~~,
 \label{scalarC3}
\eea
for  other KK modes ($n\neq 0$). 

\bigskip
\noindent
{\bf \under{Case 4}: Singlet sector II $B_z$}

The spectrum is the same as in  Case~1, but the mode functions are 
non-vanishing only in the $B_z$ part.   They  are given by
\bea
 C^B_{\vph,n} \eql \sqrt{\frac{2}{k}}
 \brc{\frac{4}{\pi^2\lmd_n^2}-F_{1,0}^2}^{-1/2}, \nonumber\\
\noalign{\kern 10pt}
 C^{\aLR{\pm}}_{\vph,n} \eql C^{\hat{a}}_{\vph,n} = 0, 
 \label{scalarC4}
\eea
where $\aLR{\pm}=1,2,3$ and $\hat{a}=1,2,3,4$. 

\bigskip

Notice that the spectrum depends on $\thH$ only in the doublet sector. 
In other words, the Higgs field couples only to the doublet sector.

\section{Spectrum and mode functions of fermions}

Masses of quarks and leptons can originate not only from gauge interactions
and bulk kink masses in the fifth dimension, but also from brane interactions
with additional fermion fields on the branes.  Indeed such additional 
interactions seem necessary to realize the observed mass spectrum and 
gauge couplings in the quark and lepton sectors.  
The main focus in the present paper is 
gauge-Higgs interactions, and we defer, to a separate paper,  detailed 
discussions about how to construct realistic models.    
At the moment we merely mention that one can 
introduce   chiral spinor fields~$\chi_{R}$  
on the Planck brane and $\chi_L$ on the TeV brane, 
which have mixing terms with the bulk fermion~$\Psi$. 
Let us take  $\eta_{0,\pi} =+1$ in (\ref{Z2parity}) so that 
the $Z_2$-parities are assigned as  Table~\ref{fermion_parity}
for a fermion multiplet in the spinor representation of $SO(5)$. 
We further suppose that $\chi_{Ri}$  ($i=1,2$) and 
$\chi_L$ have the same quantum 
number as $Q_{Ri}$ and $q_L$, where $i$ denotes the $\suR$-doublet index. 
The Lagrangian in the fermionic sector, then,  would be  
\bea
&&\hskip -1cm 
 \cL_{\rm ferm} = \sqrt{-G} ~ \bigg[ ~  i\bar{\Psi}\Gm^N\cD_N\Psi
 -iM_\Psi\vep\bar{\Psi}\Psi \cr
 \noalign{\kern 5pt}
 &&\hskip 1.9cm 
 +\sum_{i=1}^2\brc{i\bar{\chi}_{Ri}\gm^\nu\cD_\nu\chi_{Ri} 
 -\brkt{i\mu_{Qi}\bar{\chi}_{Ri}Q_{Li}+\hc}}\dlt(y)  \cr
  \noalign{\kern 5pt}
&&\hspace{19mm} 
 + \Big\{ i\bar{\chi}_L\gm^\mu\cD_\mu\chi_L
 -\brkt{i\mu_q\bar{\chi}_L q_R+\hc} \Big\}
   \dlt(y-\pi R)   ~  \bigg] ~, 
 \label{L_ferm}
\eea
where $\mu_Q$ and $\mu_q$ are brane-mass parameters 
of mass-dimension  $1/2$.   With these additional parameters 
a realistic spectrum can be reproduced.

In the subsequent discussions, however, we restrict ourselves to fermions
without brane interactions, setting $\mu_Q=\mu_q=0$ and dropping
 $\chi_{Ri}$ and $\chi_L$.  Accordingly the index $i$ of $Q_{Li}$ is
 suppressed.  
We describe below how the mass spectrum and gauge couplings are
determined, and what kind of potential problems arise in the simplified
model.

\begin{table}[t]
\caption{The $Z_2$-parities of the fermions. We take the same parity 
assignment at both boundaries.}  
\label{fermion_parity}
\begin{center}
\begin{tabular}{cccc} \hline\hline 
 \rule[-2mm]{0mm}{7mm} $q_R$ & $Q_R$ & 
 $q_L$ & $Q_L$  \\ \hline 
 even & odd & odd & even  \\ \hline 
\end{tabular}
\end{center}
\end{table}

\subsection{General solutions in the bulk}

The linearized equations of motion are 
\bea
 && e^{\sgm}\gm^\mu\der_\mu q_R-(\der_y-2\sgm'+M_\Psi\vep)q_L
 +\frac{ig_A}{2\sqrt{2}}A_y^{{\rm c}\hat{4}}Q_L  =0 ~, \nonumber\\
\noalign{\kern 5pt}
 && e^{\sgm}\gm^\mu\der_\mu Q_R-(\der_y-2\sgm'+M_\Psi\vep)Q_L
 +\frac{ig_A}{2\sqrt{2}}A_y^{{\rm c}\hat{4}}q_L = 0 ~, \nonumber\\
\noalign{\kern 5pt}
 && e^{\sgm}\gm^\mu\der_\mu q_L+(\der_y-2\sgm'-M_\Psi\vep)q_R
 -\frac{ig_A}{2\sqrt{2}}A_y^{{\rm c}\hat{4}}Q_R = 0 ~, \nonumber\\
\noalign{\kern 5pt}
 && e^{\sgm}\gm^\mu\der_\mu Q_L+(\der_y-2\sgm'-M_\Psi\vep)Q_R
 -\frac{ig_A}{2\sqrt{2}}A_y^{{\rm c}\hat{4}}q_R  = 0~.  
\label{fm_lin_eq}
\eea
From the parity assignment and  linearized equations of motion,  the boundary conditions for the bulk fermions~$(q,Q)$ are determined. 
In the conformal coordinate~$z=e^{\sgm(y)}$, they are written as 
\bea
 && D_-(c)\hat{q}_R  = 0 ~,  \hskip .8cm  \hat{q}_L = 0 ~,
 \nonumber\\
 && \hat{Q}_R = 0 ~,  \hskip 1.7cm  D_+(c)\hat{Q}_L  = 0 ~,
 \label{fm_bdcd}
\eea
at both $z=1$ and $z=\zp$.
Here $\hat{q}\equiv z^{-2}q$, $\hat{Q}\equiv z^{-2}Q$, and 
\be
 D_\pm(c) \equiv \pm\frac{d}{dz}+\frac{c}{z} ~, \label{def_Dpm}
\ee
where $c\equiv M_\Psi/k$. 
We remark that when there are brane interactions with localized fermions, 
i.e. when $\mu_q, \mu_Q \not= 0$, the above boundary conditions are
modified, becoming  no longer  Dirichlet- nor Neumann-type.

We expand the 5D fermion fields into 4D K.K. modes. 
\bea
 \hat{q}_R(x,z) \eql \sum_n\hfqR{n}(z)\psi_R^{(n)}(x), \;\;\;\;\;
 \hat{q}_L(x,z) = \sum_n\hfqL{n}(z)\psi_L^{(n)}(x), \nonumber\\
 \hat{Q}_R(x,z) \eql \sum_n\hfQR{n}(z)\psi_R^{(n)}(x),  \;\;\;\;\;
 \hat{Q}_L(x,z) = \sum_n\hfQL{n}(z)\psi_L^{(n)}(x). 
 \label{fm_md_ex1}
\eea
It follows from (\ref{fm_lin_eq}) that  equations for an eigenmode with a mass 
eigenvalue~$m_n=k\lmd_n$ are  given by
\bea
 && D_-(c)\hfqR{n}(z)-i\frac{\dot{\tht}}{2}\hfQR{n}(z) 
 = \lmd_n\hfqL{n}(z),  \nonumber\\
 && D_-(c)\hfQR{n}(z)-i\frac{\dot{\tht}}{2}\hfqR{n}(z)
  = \lmd_n\hfQL{n}(z), \nonumber\\
 && D_+(c)\hfqL{n}(z)+i\frac{\dot{\tht}}{2}\hfQL{n}(z) 
 = \lmd_n\hfqR{n}(z),  \nonumber\\
 && D_+(c)\hfQL{n}(z)+i\frac{\dot{\tht}}{2}\hfqL{n}(z) 
 = \lmd_n\hfQR{n}(z) ~.
 \label{fm_md_eq}
\eea
Here $\dot{\tht}\equiv d\tht/dz$ 
where $\tht(z)$ is defined in (\ref{def_tht}). 
The  orthonormal relations are 
\bea
&&\hskip -1cm
\int_1^{\zp}\frac{dz}{k}\brc{\brkt{\hfqR{l}}^*\hfqR{n}
 +\brkt{\hfQR{l}}^*\hfQR{n}} =\dlt_{n,l}   ~~, \cr
\noalign{\kern 10pt}
&&\hskip -1cm  
\int_1^{\zp}\frac{dz}{k}\brc{\brkt{\hfqL{l}}^*\hfqL{n}
 +\brkt{\hfQL{l}}^*\hfQL{n}} =\dlt_{n,l} ~~. 
 \label{orthonormal_fermion}
\eea

In order to solve the mode equations, it is convenient to 
move to the twisted  basis defined in (\ref{def_Omg}) and
(\ref{expr_Omg}), in which
\be
 \vct{\tl{q}}{\tl{Q}} \equiv \Omg(z)\vct{\hat{q}}{\hat{Q}} ~~. 
\ee
The mode equations  are simplified to 
\bea
 D_-(c)\fqR{n} \eql \lmd_n\fqL{n}, \;\;\;\;\;
 D_+(c)\fqL{n} = \lmd_n\fqR{n}, \nonumber\\
 D_-(c)\fQR{n} \eql \lmd_n\fQL{n},  \;\;\;\;\;
 D_+(c)\fQL{n} = \lmd_n\fqR{n},  \label{fm_md_eq2}
\eea
where 
\be
 \vct{\fqR{n}}{\fQR{n}} \equiv \mtrx{\cth}{i\sth}{i\sth}{\cth}
 \vct{\hfqR{n}}{\hfQR{n}}, \;\;\;\;\;
 \vct{\fqL{n}}{\fQL{n}} \equiv \mtrx{\cth}{i\sth}{i\sth}{\cth}
 \vct{\hfqL{n}}{\hfQL{n}}. 
\ee
Then (\ref{fm_md_ex1}) becomes 
\bea
 \tl{q}_R(x,z) \eql \sum_n\fqR{n}(z)\psi_R^{(n)}(x), \;\;\;\;\;
 \tl{q}_L(x,z) = \sum_n\fqL{n}(z)\psi_L^{(n)}(x), \nonumber\\
 \tl{Q}_R(x,z) \eql \sum_n\fQR{n}(z)\psi_R^{(n)}(x),  \;\;\;\;\;
 \tl{Q}_L(x,z) = \sum_n\fQL{n}(z)\psi_L^{(n)}(x). 
 \label{fm_md_ex2}
\eea
The general solutions of Eqs.(\ref{fm_md_eq2}) are 
\bea
 \fqR{n}(z) \eql z^{\frac{1}{2}}\brc{a^q_{n} J_{\alp-1}(\lmd_n z) 
 +b^q_{n} Y_{\alp-1}(\lmd_n z)}, \nonumber\\
 \fQR{n}(z) \eql z^{\frac{1}{2}}\brc{a^Q_{n} J_{\alp-1}(\lmd_n z) 
 +b^Q_{n} Y_{\alp-1}(\lmd_n z)}, \nonumber\\
 \fqL{n}(z) \eql z^{\frac{1}{2}}\brc{a^q_{n} J_\alp(\lmd_n z) 
 +b^q_{n} Y_\alp(\lmd_n z)}, \nonumber\\
 \fQL{n}(z) \eql z^{\frac{1}{2}}\brc{a^Q_{n} J_\alp(\lmd_n z) 
 +b^Q_{n} Y_\alp(\lmd_n z)}, 
 \label{fm_gen_sol}
\eea
where $\alp\equiv (M_\Psi/k)+\frac{1}{2}$. 
The eigenvalues~$\lmd_n$ and the coefficients~$a_n$'s and $b_n$'s are determined 
by the boundary conditions~(\ref{fm_bdcd}).

\subsection{Mass eigenvalues and mode functions}

From the conditions~(\ref{fm_bdcd}) at $z= z_\pi$, the ratios between 
$a_n$'s and $b_n$'s 
in (\ref{fm_gen_sol}) are determined so that the mode functions are 
written by using the function~$F_{\alp,\bt}(u,v)$ 
defined in (\ref{def_Fs}) as 
\bea
 \fqR{n}(z) \eql C^q_n z^{\frac{1}{2}}
 F_{\alp-1,\alp}(\lmd_n z,\lmd_n\zp) ~, \nonumber\\
 \fQR{n}(z) \eql C^Q_n z^{\frac{1}{2}}
 F_{\alp-1,\alp-1}(\lmd_n z,\lmd_n\zp) ~, \nonumber\\
 \fqL{n}(z) \eql C^q_n z^{\frac{1}{2}}
 F_{\alp,\alp}(\lmd_n z,\lmd_n\zp) ~, \nonumber\\
 \fQL{n}(z) \eql C^Q_n z^{\frac{1}{2}}
 F_{\alp,\alp-1}(\lmd_n z,\lmd_n\zp) ~. 
 \label{fm_sol2}
\eea
The eigenvalue $\lambda_n$ and constants~$C^q_n$ and $C^Q_n$ are determined 
by the remaining boundary conditions in (\ref{fm_bdcd}) at $z=1$. 
By making use of  (\ref{fm_sol2}), the two conditions for right-handed components 
in (\ref{fm_bdcd}) at $z=z_\pi$ are rewritten as~\footnote{
The remaining conditions in (\ref{fm_bdcd})  at $z=z_\pi$ provide the same conditions 
on $(C^q_n,C^Q_n)$ as (\ref{eq_for_CLs}).}
\be
 \mtrx{\cw F_{\alp,\alp}}{-i\sw F_{\alp,\alp-1}}
 {-i\sw F_{\alp-1,\alp}}{\cw F_{\alp-1,\alp-1}}
 \vct{C^q_n}{C^Q_n} = 0  ~~.
 \label{eq_for_CLs}
\ee
Here $F_{\alpha ,\beta} = F_{\alpha , \beta} (\lmd_n,\lmd_n\zp)$. 
For a nontrivial solution to exist, the determinant of the $2\times 2$ matrix 
in Eq.(\ref{eq_for_CLs}) must vanish, which leads to 
\be
 \pi^2\lmd_n^2\zp F_{\alp-1,\alp-1} F_{\alp,\alp} = 4\sw^2 ~~. 
 \label{det_Mf}
\ee
We have used the last formula in (\ref{propty_Fs}). 
This is the equation that determines the mass spectrum~$\brc{\lmd_n}$. 
Once $\lmd_n$ is determined , the corresponding $C^q_n$ and $C^Q_n$ 
are fixed by (\ref{eq_for_CLs}) with the normalization 
condition~(\ref{orthonormal_fermion}). 
The result is 
\bea
 C^q_n \eql i\tw\frac{F_{\alp,\alp-1}}{F_{\alp,\alp}}C^Q_n, 
 \nonumber\\
 \noalign{\kern 10pt}
 C^Q_n \eql \sqrt{2k}  \bigg[  ~  \frac{4}{\pi^2\lmd_n^2}
 +\frac{\pi^2\lmd_n^2\zp^2}{4\sw^2\cw^2}
 F_{\alp,\alp-1}^2 F_{\alp-1,\alp-1}^2  
  -\frac{F_{\alp,\alp-1}^2}{\cw^2} 
  -\frac{ F_{\alp-1,\alp-1}^2}{\sw^2} \bigg]^{-1/2}. 
 \label{values_CL}
\eea

\section{Masses and wave functions of light particles} \label{ap_mass}

Approximate expressions of the
masses and wave functions of light particles such as $W$, $Z$, 
quarks and leptons can be obtained.  The mass of the 
4D Higgs particle is generated at the quantum level.  Its mass is 
estimated from the effective potential for the Yang-Mills AB phase $\thH$.

\subsection{Gauge sector}

\begin{figure}[t,b]
\centerline{\includegraphics[width=5cm]{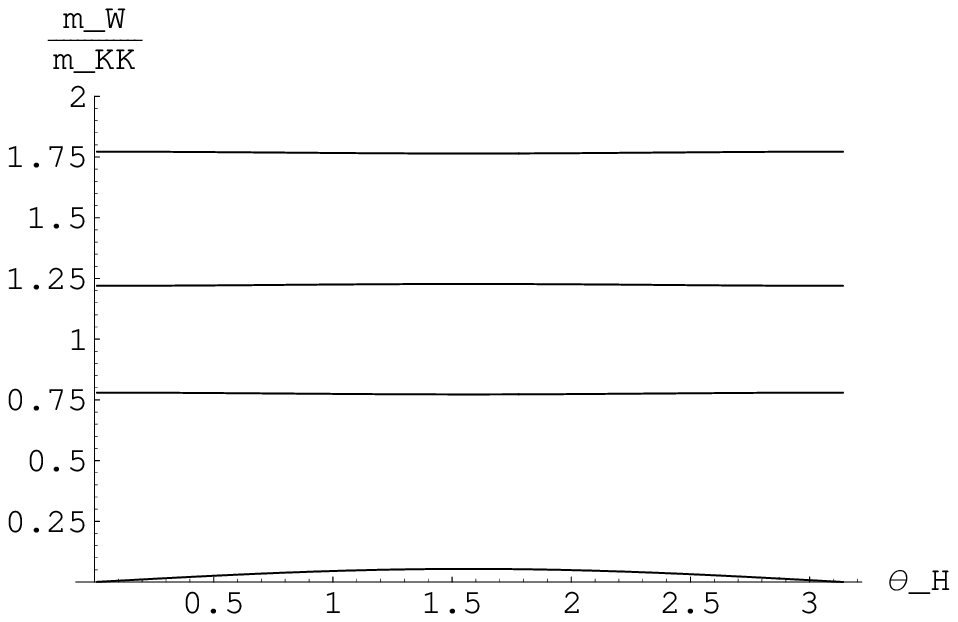}
\hskip .2cm
\includegraphics[width=5cm]{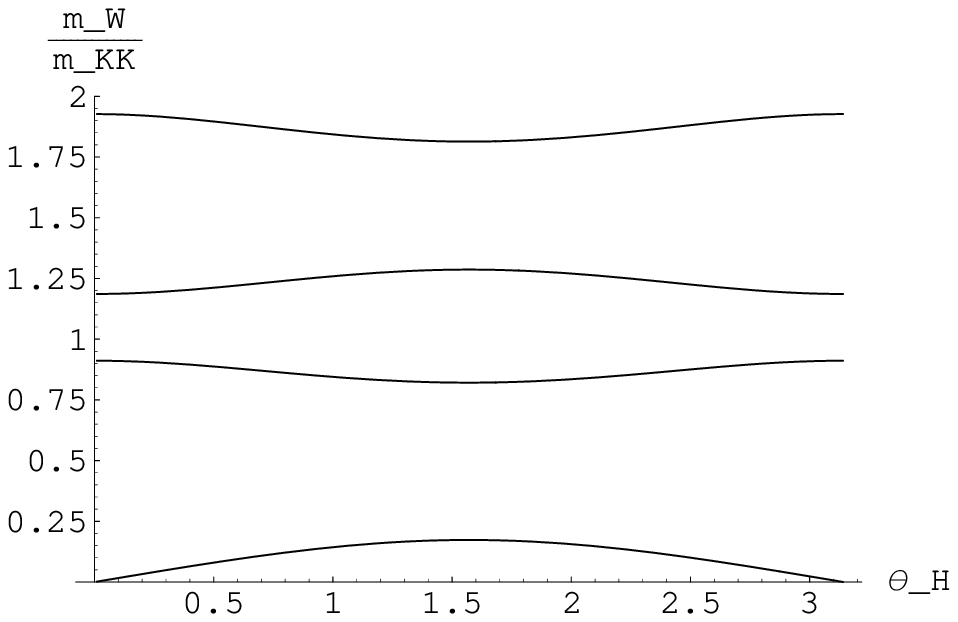}
\hskip .2cm
\includegraphics[width=5cm]{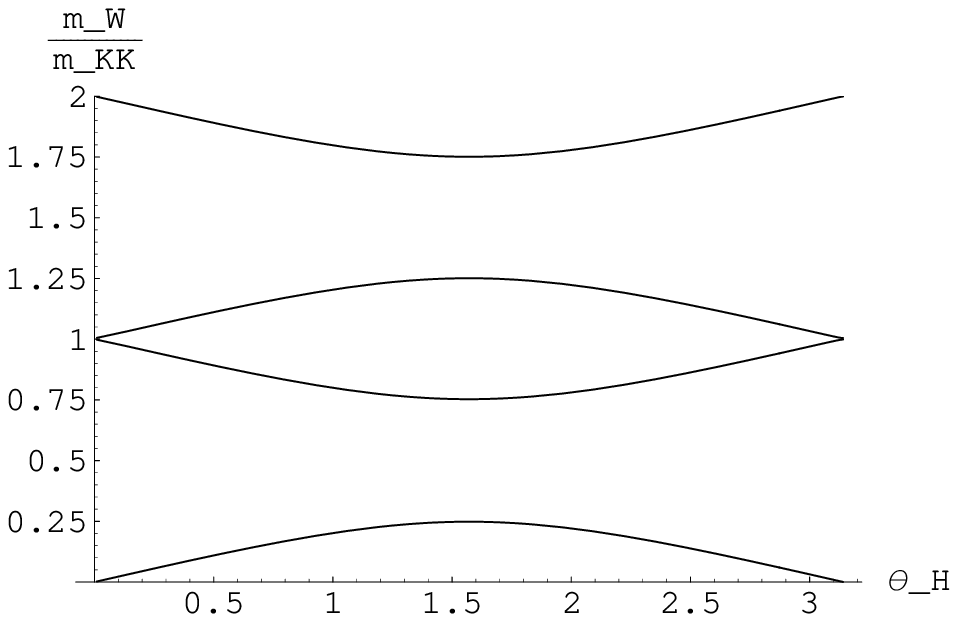}}
\caption{The masses of $W$ boson and its K.K. excited states
are depicted in the unit of $\mKK$ as functions of $\thw$
for $k\pi R=35, 3.5$, and $0.35$ from the left to the right.
}
\label{Wmass_thH}
\end{figure}

The masses and wave functions of the $W$ and $Z$ bosons have 
nontrivial $\thH$ dependence.  
They belong to  the case 2 of the charged sector 
in Sec.~\ref{charged_sector} and 
the case 3 of the neutral sector in Sec.~\ref{neutral_sector} 
whose  mass spectra are determined by 
\bea
 \pi^2\lmd_n^2\zp F_{0,0}F_{1,1} \eql 2\sin^2\thH ~~, \cr
\noalign{\kern 5pt}
 \pi^2\lmd_n^2\zp F_{0,0}F_{1,1} \eql 2(1+\sph^2)\sin^2\thH ~~, 
 \label{mW_determination}
\eea
respectively.  Here $F_{\alpha, \beta} = F_{\alpha, \beta}(\lmd_n,\lmd_n\zp)$.
These equations are similar to those in the $SU(3)$ model discussed in 
Ref.~\citen{HNSS},  but there is an important difference
in the  numerical factors on the right sides of the two equations above.
In the $SU(3)$ model one has $4 \sin^2 \onehalf \thH$ in place of
$2 \sin^2 \thH$ in the equation determining the $W$ 
mass.\footnote{
The masses of the $Z$ boson and its K.K. modes were not discussed
in Ref.~\citen{HNSS} as the correct value of the Weinberg angle~$\thw$ 
is not obtained  in the $SU(3)$ model. }
Fig.~\ref{Wmass_thH} depicts the masses of the KK tower of the $W$ boson 
as functions of $\thH$ for $k\pi R=35,3.5,0.35$. 
Due to the numerical factor mentioned above,
the mass spectrum does not approach 
a linear spectrum in $\thH$ in the flat limit ($k\pi R\to 0$) 
in contrast to the $SU(3)$ model. 
(See Fig.~1 in Ref.~\citen{HNSS}.)
For $0 < k\pi R\ll 1$,  
\be
 \lmd_n\zp \simgt \lmd_n \simgt \frac{\mKK}{k}=\frac{\pi}{\zp-1} 
 \simeq \frac{\pi}{k\pi R} \gg 1 \hskip 1cm  (n\geq 1)~.
\ee
With  (\ref{asymp_bhv}),  the left-hand side of (\ref{mW_determination}) becomes 
\be
\pi^2  \lmd_n^2\zp F_{0,0}F_{1,1} \simeq 4\sin^2 (\pi R m_n) 
\hskip 1cm (n\geq 1) ~. 
\ee
The mass eigenvalues~$m_n=k\lmd_n$ become linear functions of $\thH$
in the flat limit  only if the numerical factor in the right-hand side of 
Eq.\ (\ref{mW_determination}) is $4$.\footnote{
When the numerical factor is 4, it can be easily shown that 
the mass of the lightest ($n=0$) mode also becomes a linear function of $\thH$ 
in the flat limit. 
} 
Therefore the K.K. level-crossing does not occur as $\thH$ increases 
from 0 to $\pi$ in our model even in the flat geometry. 
This is one of the distinctive properties of the $SO(5)\times\ubl$ model. 

As can be seen from Fig.~\ref{Wmass_thH}, the mass of the lightest mode 
is much lighter than the K.K. mass scale~$\mKK$ 
when the warp factor~$\zp=e^{k\pi R}$ is large, that is,  
when $\lmd_0,\lmd_0\zp\ll 1$. 
Thus one may use (\ref{asymp_F}) 
to obtain an approximate expression for the mass of the lightest mode
in each category  
determined by (\ref{mW_determination}) and the corresponding mode function 
from (\ref{charged_Cs2}) or (\ref{neutral_Cs3}). 
For the $W$ boson the mass is given by 
\be
 m_W = \frac{\mKK}{\pi}\sqrt{\frac{1}{k\pi R}} ~ \abs{\sin\thH} 
 \bigg\{ 1 + \cO \Big( \frac{\pi^2m_W^2}{\mKK^2} \Big) \bigg\}
 \label{ap_mW}
\ee
with its mode function 
\be
 \tl{h}^{\chL}_{A,0}(z) \simeq \frac{1+\cos\thH}{2\sqrt{\pi R}} ~~,~~
 \tl{h}^{\chR}_{A,0}(z) \simeq \frac{1-\cos\thH}{2\sqrt{\pi R}} ~~,~~
 \tl{h}^{\hat{\pm}}_{A,0}(z) \simeq - \frac{\sin\thH}{\sqrt{2\pi R}}
 \brkt{1 - \frac{z^2}{\zp^2}} ~~. 
 \label{ap_Wmf}
\ee
For the $Z$ boson the mass is given by
\be
 m_Z = \frac{\mKK}{\pi}\sqrt{\frac{1+\sph^2}{k\pi R}} ~ \abs{\sin\thH} 
 \bigg\{ 1 + \cO \Big( \frac{\pi^2m_Z^2}{\mKK^2} \Big) \bigg\}
 \label{ap_mZ}
\ee
with its mode function
\bea
&&\hskip -1cm
 \tl{h}^{\nL}_{A,0}(z) \simeq \frac{\cph^2+\cos\thH(1+\sph^2)}
 {2\sqrt{(1+\sph^2)\pi R}} ~~,~~
 \tl{h}^{\nR}_{A,0}(z) \simeq \frac{\cph^2-\cos\thH(1+\sph^2)}
 {2\sqrt{(1+\sph^2)\pi R}} ~~, \nonumber\\
&&\hskip -1cm
 \tl{h}^{\hat{3}}_{A,0}(z) \simeq  - \sin\thH\sqrt{\frac{1+\sph^2}{2\pi R}}
 \brkt{ 1-  \frac{z^2}{\zp^2}} ~~,~~
 \tl{h}^B_{A,0}(z) \simeq -\frac{\sph\cph}{\sqrt{(1+\sph^2)\pi R}} ~~. 
 \label{ap_Zmf}
\eea

We stress that $m_W$ and $m_Z$ are not proportional to the VEV of 
the 4D Higgs field,  or $\thH$, in contrast to the ordinary Higgs mechanism
in four dimensions.   The mechanism of mass generation for the 4D gauge bosons 
in the gauge-Higgs unification scenario involves not only 4D gauge fields
and scalar fields in each KK level, but also fields in other KK levels. 
The lowest mode in each KK tower necessarily mixes with heavy K.K. modes 
when $\thH$ acquires a nonzero value.   Furthermore, there is  mixing with
other components of the gauge group at non-vanishing $\thH$ as well.  
The spectrum and mixing is such that they become periodic in $\thH$
with a period $2\pi$.

From (\ref{ap_mW}) and (\ref{ap_mZ}) the Weinberg angle~$\thw$ 
determined from $m_W$ and $m_Z$ becomes
\beqn
&&\hskip -1cm
 \sin^2\thw \equiv 1-\frac{m_W^2}{m_Z^2}    \cr
\noalign{\kern 5pt}
&&\hskip .4cm
\simeq \frac{\sph^2}{1+\sph^2} = \frac{g_B^2}{g_A^2+2g_B^2}
= \frac{g_Y^2}{g_A^2 + g_Y^2} ~~. 
 \label{thW}
\eeqn
The approximate equality in the second line is valid 
to the   $\cO(0.1\%)$ accuracy for $\mKK=\cO(\mbox{TeV})$. 
In the last equality   the relation  $g_Y = g_A g_B / \sqrt{g_A^2 + g_B^2}$
has been made use of.  
Note that $\sph$ defined in (\ref{def_vph}) satisfies $\sph \simeq \tan\thw$.
The Weinberg angle $\thw$ may be determined 
from the vertices in the neutral current interactions. 
As we will see in   (\ref{gZL})  below, 
$\thw$ determined this way  coincides 
with that in (\ref{thW}) to good accuracy. 
Thus the $\rho$ parameter is nearly one in our model. 
We remark that the $\rho$ parameter substantially deviates from 1
in the flat limit ($\pi kR \go 0$) when $\thH$ is nonvanishing.

According to the classification in the gauge sector, 
it is convenient to devide the K.K. modes of the charged sector into two classes: 
$(\tl{W}^{(n)}_\mu,W^{(n)}_\mu)$,   corresponding 
to the cases~1 and 2  in the charged sector.
Similarly the K.K. modes of the neutral sector are divided  into three classes: 
$(A^{\gm(n)}_\mu,\tl{Z}^{(n)}_\mu,Z^{(n)}_\mu)$,  corresponding 
to the cases~1, 2,  and 3 in the neutral sector. 
The $W$ and $Z$ bosons are $W^{(0)}_\mu$ and $Z^{(0)}_\mu$, 
respectively.

\subsection{Fermion sector} \label{Ap_formulae_fermion}

The mass spectrum in the fermion sector is determined by Eq.(\ref{det_Mf}). 
For  $\lmd_0\zp\ll 1$ we obtain from (\ref{det_Mf}) and (\ref{asymp_F}) 
\cite{HNSS,Carena}
\be
 m_0 = k\lmd_0 \simeq k\brkt{\frac{\alp(\alp-1)}{\zp\sinh(\alp k\pi R)
 \sinh((\alp-1)k\pi R)}}^{1/2}\abs{\sw}. 
 \label{ap_m01}
\ee
The mode functions for the lightest mode~$\psi^{(0)}$ are obtained 
from Eq.(\ref{eq_for_CLs}) 
with the normalization condition~(\ref{orthonormal_fermion}). 
For $\alp>1$, for example, they are approximately expressed as~\footnote{
The subleading terms in $\fqL{0}(z)$ and $ \fQR{0}(z)$ dropped in 
(\ref{ap_fmf1})  become comparable to the leading terms at $z=\zp$,  
but they remain suppressed in the bulk. 
} 
\bea
&&\hskip -1cm
 \fqR{0}(z) \simeq -i\sqrt{2k\alp}\zp^{-\alp}z^{\alp-\frac{1}{2}}
 ~~, ~~
 \fQR{0}(z) \simeq \sw\cw\sqrt{2k\alp}\zp^{-\alp}z^{\frac{3}{2}-\alp} ~~, 
 \nonumber\\
\noalign{\kern 10pt}
&&\hskip -1cm
\fqL{0}(z) \simeq i\sw\sqrt{2k(\alp-1)}z^{\frac{1}{2}-\alp}
~~,~~  
 \fQL{0}(z) \simeq \cw\sqrt{2k(\alp-1)}z^{\frac{1}{2}-\alp} ~~. 
 \label{ap_fmf1}
\eea
It follows that the right-handed component~$\psi_R^{(0)}$ is 
localized around the TeV brane while the left-handed component~$\psi_L^{(0)}$ 
 around the Planck brane.\footnote{
$\fQR{0}(z)$ can be localized at the Planck brane if $\alp>3/2$, but 
it is exponentially suppressed compared to 
$\fqR{0}(z)$ for $z>1$ in such a case. } 
For $\alp<0$, on the other hand, $\psi_R^{(0)}$ is localized 
around the Planck brane while $\psi_L^{(0)}$  around the TeV brane. 

With (\ref{ap_m01}) the hierarchical fermion mass spectrum can be reproduced
by choosing $\alp$ in an $\cO(1)$ range~\cite{HNSS}. 
However there is a serious problem in this senario as pointed out 
in Ref.~\citen{SH1}. 
The gauge couplings to the $W$ and  $Z$ bosons deviate from 
the observed values for at least one chiral component of the fermion 
when $\thH$ is nonzero.     This is due to the fact that 
the wave function of one of the chiral components is inevitably 
localized around the TeV brane where the mode functions for the $W$ and $Z$ 
bosons deviate from the constant values 
(see (\ref{ap_Wmf}) and (\ref{ap_Zmf})). 

This problem can be  avoided by introducing   boundary fields and turning on   
boundary mass terms with bulk fermions as described with the action (\ref{L_ferm}).  
There are two origins for fermion masses;  the Yang-Mills AB phase~$\thH$ 
and  boundary masses~$\mu_q$, $\mu_Q$. 
It can be shown that for $\alp>1$ and 
$z_\pi^{2(1-\alpha)} \ll  ( \mu_Q^2/k) \zp^2  \ll \mu_q^2/k  \simeq \cO(1)$
the lightest mass becomes $m_0 \simeq  \sqrt{k(\alp-1)}\mu_Q$ and
its wave functions are suppresed at the TeV brane for both $L$ and $R$
chiral components.  
Further  $\suR$ is broken at the Planck brane so that
$\mu_Q$ can be chosen to be different for the upper and the lower 
components of $Q_L$.  Thus one can realize the observed fermion masses 
with appropriate $\mu_Q$.   The detailed discussions will be given in a 
separate paper.

\section{Gauge-Higgs  self-couplings} \label{couplings}

Once wave functions of the $W$ and $Z$ bosons and the Higgs boson are
determined, the effective four-dimensional couplings among them can be
calculated as overlap integrals in the fifth dimension.   As $\thH$ becomes
nonvanishing,  the form of wave functions substantially changes with large mixing
so that the effective four-dimensional couplings are expected to have nontrivial 
$\thH$-dependence in general.    This behavior of the couplings provides
crucial tests for the gauge-Higgs unification scenario.

The $WWZ$ coupling has been indirectly measured in the LEP2 experiment of
$W$ pair production.  The standard model fits the data well within a few percents 
so that  deviation from the value in the standard mode could rule out the model under 
consideration.  As we shall see below,  the $WWZ$ coupling in the gauge-Higgs
unification in the Randall-Sundrum warped spacetime remains almost universal
as in the standard model.  In the gauge-Higgs unification in flat spacetime, however,
substantial deviation results.  

Important predictions from the gauge-Higgs unification scenario are obtained 
for the $WWH$, $ZZH$, $WWHH$ and $ZZHH$ couplings where $H$ stands for
the Higgs boson.  These couplings are
suppressed compared with those in the standard model.  This gives a crucial
test to be performed at LHC in the coming years.

\subsection{$WWZ$ coupling}

The self-couplings among $W$, $Z$ and Higgs bosons  are determined from
the interaction terms in the twisted basis
\beqn
&&\hskip -1cm
\int_{z_0}^{z_\pi}  \frac{dz}{kz}  \, 
\bigg\{ ig_A    \eta^{\mu\rho} \eta^{\nu\sigma} 
    \tr (\dd_\mu \tilde A_\nu - \dd_\nu \tilde A_\mu)   [\tilde A_\rho , \tilde A_\sigma ]
+ 2 i g_A k^2 \eta^{\mu\nu} \tr (\dd_\mu \tilde A_z - \dd_z \tilde A_\mu)  
       [\tilde A_\nu , \tilde A_z ]  \cr
\noalign{\kern 10pt}
&&\hskip 1.5cm 
+ \frac{1}{2} g_A^2    \eta^{\mu\rho} \eta^{\nu\sigma} 
 \tr  [\tilde A_\mu , \tilde A_\nu ]  [\tilde A_\rho , \tilde A_\sigma ]
 + g_A^2 k^2 \eta^{\mu\nu} \tr [\tilde A_\mu , \tilde A_z ]  [\tilde A_\nu , \tilde A_z ]     ~ \bigg\}
 \label{action5}
\eeqn
by inserting the wave functions of $W$, $Z$ and $H$.  
The relevant part of the expansion of the gauge fields  is
\beqn
&&\hskip -1cm
\tilde A_\mu = W^{(0)}_\mu(x) \Big\{ \tilde h^{\chL}_{W}(z) T^{-_{\rm L}}
+  \tilde h^{\chR}_{W}(z) T^{-_{\rm R}} 
+  \tilde h^{\hat \pm}_{W}(z) T^{\hat -} \Big\} \cr
\noalign{\kern 5pt}
&&\hskip 0.cm
+  W^{(0) \dagger}_\mu(x) \Big\{ \tilde h^{\chL}_{W}(z) T^{+_{\rm L}}
+  \tilde h^{\chR}_{W}(z) T^{+_{\rm R}} 
+  \tilde h^{\hat \pm}_{W}(z) T^{\hat +} \Big\} \cr
\noalign{\kern 5pt}
&&\hskip 0.cm
+ Z^{(0)}_\mu(x) \Big\{ \tilde h^{\nL}_{Z}(z) T^{\nL}
+  \tilde h^{\nR}_{Z}(z) T^{\nR} +  \tilde h^{\hat 3}_{Z}(z) T^{\hat 3} \Big\} ~~,  \cr
\noalign{\kern 5pt}
&&\hskip -1cm
\tilde A_z = H^{(0)}(x)  \tilde h^{\hat 4}_{H}(z) T^{\hat 4} ~~.
\label{gauge_expansion1}
\eeqn
Here $T^{\chL} = (T^{1_{\rm L}} \pm i T^{2_{\rm L}} )/\sqrt{2}$ etc..
The $W$ wave functions
$\tl{h}^{{\pm_{\rm R,L}}}_{W}(z) =\tl{h}^{{\pm_{\rm R,L}}}_{A,0}(z)$ and  
$\tl{h}^{\hat{\pm}}_{W} (z) = \tl{h}^{\hat{\pm}}_{A,0} (z)$
are given by  (\ref{sol2}) with (\ref{charged_Cs2}).
The $Z$ wave functions 
$\tl{h}^{3_{\rm R,L}}_{Z}(z)= \tl{h}^{3_{\rm R,L}}_{A,0}(z)$ and  
$\tl{h}^{\hat3}_{Z} (z) = \tl{h}^{\hat3}_{A,0} (z)$
are given by  (\ref{sol2}) with (\ref{neutral_Cs3}).
The Higgs wave function $\tl{h}^{\hat 4}_{H}(z)= \tl{h}^{\hat 4}_{\vph,0}(z)$ is 
given by (\ref{scalar_wave3}).  

\ignore{
Note that $ \tilde h^{+_L}_{A,0}(z) =  \tilde h^{-_L}_{A,0}(z) 
\equiv  \tilde h^{\chL}_{A,0}(z)$ etc.   
In (\ref{gauge_expansion1}),  $ \tilde h^{\chL}_{A,0}$,  $ \tilde h^{\chR}_{A,0}$,
and $ \tilde h^{\hat \pm}_{A,0}$ are those of $W$ bosons, whereas 
$\tilde h^{3_L}_{A,0}$, $\tilde h^{3_R}_{A,0}$,  and $\tilde h^{\hat 3}_{A,0}$ are
those of $Z$ bosons.  }

The  $WWZ$ coupling is evaluated by inserting (\ref{gauge_expansion1}) 
into  the first term in (\ref{action5}) and integrating over $z$.  The result is
\bea
 \cL^{(4)}_{WWZ} \eql 
 ig_{WWZ} \Big\{ 
( \der_\mu W^{(0) \dagger}_\nu-\der_\nu W^{(0) \dagger}_\mu )
 W^{(0)\mu}Z^{(0)\nu}
 -  ( \der_\mu W^{(0)}_\nu-\der_\nu W^{(0)}_\mu )
 W^{(0) \dagger \mu}Z^{(0)\nu}   \cr
 \noalign{\kern 10pt}
 &&\hskip 2cm + W^{(0)\dagger}_\mu W^{(0)}_\nu
 \brkt{\der^\mu Z^{(0)\nu}-\der^\nu Z^{(0)\mu}}  \Big\}
\eea
where the coupling $g_{WWZ}$is expressed as
overlap integrals 
\bea
 g_{WWZ} \eql  
 g_A\int_1^{\zp}\frac{dz}{kz} ~
 \bigg[   \tl{h}^{\nL}_{Z}\brc{\brkt{\tl{h}^{\chL}_{W}}^2
 +\frac{1}{2}\brkt{\tl{h}^{\hat{\pm}}_{W}}^2}
 +\tl{h}^{\nR}_{Z}\brc{\brkt{\tl{h}^{\chR}_{W}}^2
 +\frac{1}{2}\brkt{\tl{h}^{\hat{\pm}}_{W}}^2}   \cr
 \noalign{\kern 10pt}
 &&\hspace{40mm} 
 +\tl{h}^{\hat{3}}_{Z} ~\tl{h}^{\hat{\pm}}_{W}
 \brkt{\tl{h}^{\chL}_{W}+\tl{h}^{\chR}_{W}}   \bigg] ~~. 
 \label{def_gWWZ}
\eea

Note that given  $\thH$ and $k\pi R$, $m_W$ determines $k$.   $k=O(M_{pl})$ 
cooresponds to $k\pi R \sim 35$.  With these parameters we have
exact wave functions as
summarized in (\ref{sol2}),  (\ref{charged_Cs2}) and (\ref{neutral_Cs3}).
When $k\pi R \sim 35$ and the warp factor is large $e^{k\pi R} \gg 1$, the
approximate expressions for the wave functions of $W$ and $Z$ given by
(\ref{ap_Wmf}) and (\ref{ap_Zmf}) can be employed to find
\be
 g_{WWZ}  \simeq \frac{g_A}{\sqrt{(1+\sph^2)\pi R}}
 \simeq g\cos\thw  ~~. 
 \label{ap_gWWZ}
\ee
Here a dimensionless coupling $g$ is defined as 
\be
 g \equiv \frac{g_A}{\sqrt{\pi R}} ~~,  \label{def_g}
\ee
which is the 4D $\suL$ gauge coupling at $\thH=0$. 
In the last equality in (\ref{ap_gWWZ}), (\ref{thW})  has been made use of. 
We have neglected corrections suppressed by 
a factor of $(k\pi R)^{-1}\simeq 1/35$ in conformity 
with the approximation employed in deriving Eqs.(\ref{ap_mW})-(\ref{ap_Zmf}). 

These couplings (\ref{ap_gWWZ}) have the same values as those in the standard model. 
Although the wave functions vary substantially as $\thH$, the triple
gauge coupling $WWZ$ remains almost universal.  The statement can be strengthened by numerically integrating (\ref{def_gWWZ}) with
exact wave functions.  In Table~\ref{values_gWWZ}  the values of the ratio of 
the trilinear couplings  (\ref{def_gWWZ}) 
to $g_A/\sqrt{(1+\sph^2)\pi R}$ are shown for various values of 
$\thH$ and $k \pi R$.   It is clearly seen that the relation (\ref{ap_gWWZ}) 
hold to extreme accuracy when the warp factor $e^{k\pi R}$ is large.
This is very important
in the light of the experimental fact that $W$ pair production rate measured
at LEP2 is consistent with the  $WWZ$ coupling in the standard model.
In the flat spacetime limit ($k \pi R \go 0$), however, 
substantial deviation from the 
standard model results with moderate $\thH$.  When $\thH = O(1)$
gauge-Higgs unification in flat spacetime  contradicts with the data for
the $WWZ$ coupling.

\begin{table}[t, b]
\caption{The values of a ratio of $g^{(1)}_{WWZ}$($=g^{(2)}_{WWZ}$) 
to $g_A/\sqrt{(1+\sph^2)\pi R}$ for $\thH=\pi/10,\pi/4,\pi/2$ 
and $k\pi R=35,3.5,0.35$. }  
\label{values_gWWZ}
\begin{center}
\begin{tabular}{c|ccc}
 \hline\hline 
 \backslashbox{$k\pi R$}{$\thH$} & $\pi/10$ & $\pi/4$ & $\pi/2$ 
 \\ \hline 
 35 & 0.9999987 & 0.999964 & 0.99985 \\ \hline 
 3.5 & 0.9999078 & 0.996993 & 0.98460 \\ \hline
 0.35 & 0.9994990 & 0.979458  & 0.83378 \\ \hline 
 \end{tabular}
\end{center}
\end{table}

As can be seen in the overlap integrals in (\ref{def_gWWZ}), weight of
wave functions is best measured in the $y = k^{-1} \ln z$ coordinate.
The wave functions of the gauge bosons in the warped spacetime in the $y$
coordinate remain almost constants except in a tiny region near $y=\pi R$
even at $\thH \not= 0$.  (See (\ref{ap_Wmf}) and (\ref{ap_Zmf}).)
In the flat spacetime, however, wave functions
are deformed substantially at  $\thH \not= 0$ to have nontrivial
$y$-dependence in the entire region, which contributes to substantial
deviation in the $WWZ$ coupling from the standard model.

\subsection{$WWH$ and $ZZH$ couplings}  \label{WWHcp}

A striking prediction of the gauge-Higgs unification scenario is obtained
for the Higgs couplings to $W$ and $Z$.  Let us consider 
the $WWH$ and $ZZH$ couplings.  The 4D Higgs filed $H = \vphi^{(0)}$
is a part of the 5D gauge fields so that the Higgs couplings to $W$ and $Z$
are completely determined by the gauge principle once $\thH$ is given.
Unlike the $WWZ$ coupling, substantial deviation from the standard
model is predicted.  

Indeed, by inserting (\ref{gauge_expansion1})  into 
the second term in (\ref{action5}),  one finds that
\be
 \cL^{(4)} = \lmd_{WWH} ~ H^{(0)}W^{(0)\mu\,\dagger}W^{(0)}_\mu 
 +  \half  \lmd_{ZZH} ~ H^{(0)}Z^{(0)\mu}Z^{(0)}_\mu+\cdots, 
\ee
where 
\bea
&&\hskip -1cm
\lmd_{WWH} = g_A k\int_1^{\zp}\frac{dz}{z}\; \tl{h}^{\hat 4}_{H}
 \brc{\tl{h}^{\hat{\pm}}_{W}\der_z
         \brkt{\tl{h}^{\chL}_{W}  -\tl{h}^{\chR}_{W}}
 -\der_z\tl{h}^{\hat{\pm}}_{W}
         \brkt{\tl{h}^{\chL}_{W}   -\tl{h}^{\chR}_{W}}}, \cr
\noalign{\kern 10pt}
&&\hskip -1cm
\lmd_{ZZH} = g_A k\int_1^{\zp}\frac{dz}{z}\; \tl{h}^{\hat 4}_{H}
 \brc{\tl{h}^{\hat{3}}_{Z}\der_z
       \brkt{\tl{h}^{\nL}_{Z}   -\tl{h}^{\nR}_{Z}}
 -\der_z\tl{h}^{\hat{3}}_{Z}
       \brkt{\tl{h}^{\nL}_{Z}   -\tl{h}^{\nR}_{Z}}}. 
\label{def_WWH}
\eea
Recall that the wave function of the Higgs field, 
$\tl{h}^{\hat 4}_{H}(z)  \propto z=e^{ky}$, is localized near the TeV brane
at $z=z_\pi$ when evaluated in the $y$-coordinate relevant 
in the integrals (\ref{def_WWH}).  The behavior of the wave functions of $W$ and $Z$ bosons 
near the TeV brane sensitively depends on $\thH$ so that nontrivial
$\thH$ dependence is expected for the $WWH$ and $ZZH$ couplings.

The integrals in (\ref{def_WWH}) can be evaluated  in a closed form.  
Consider $\lambda_{WWH}$.
Inserting the wave functions into (\ref{def_WWH})    and making use of the identity
$\dd_z \big\{ z F_{1,\beta} (\lmd z, \lmd\zp) \big\}
= \lambda z  F_{0,\beta} (\lmd z, \lmd\zp) $ and
the last relation in (\ref{propty_Fs}),  one finds 
\beqn
 \lmd_{WWH} \eql
\frac{4 g_A k^2   C^{\hat \pm}_{W} 
(C^{\pm_{\rm L}}_{W} - C^{\pm_{\rm R}}_{W})}{\pi^2 m_W z_\pi}
\int_1^{z_\pi} dz  ~  \tl{h}^{\hat 4}_{H} (z)  \cr
\noalign{\kern 10pt}
\eql  \frac{2 g_A k\sqrt{ 2k (z_\pi^2 -1) }  }{\pi^2 m_W z_\pi}   
~ C^{\hat \pm}_{W} 
(C^{\pm_{\rm L}}_{W} - C^{\pm_{\rm R}}_{W})  ~~.
\label{lamWWH1}
\eeqn
The coefficients $C^{\pm_{\rm L,R}}_{W} = C^{\pm_{\rm L,R}}_{A,0}$ and 
$C^{\hat \pm}_{W} = C^{\hat \pm}_{A,0}$ are given by
(\ref{charged_Cs2}).
Similarly for $\lambda_{ZZH}$ one finds 
\beeq
 \lmd_{ZZH} =
 \frac{2 g_A k \sqrt{ 2k (z_\pi^2 -1) }  }{\pi^2 m_Z z_\pi}   
~ C^{\hat 3}_{Z}  (C^{3_{\rm L}}_{Z} - C^{3_{\rm R}}_{Z})  ~~,
\label{lamZZH1}
\eneq
where 
the coefficients $C^{3_{\rm L,R}}_{Z} = C^{3_{\rm L,R}}_{A,0}$ and 
$C^{\hat 3}_{Z} = C^{\hat 3}_{A,0}$ are given by
 (\ref{neutral_Cs3}).

The formulas (\ref{lamWWH1}) and (\ref{lamZZH1}) are exact.
They are fairly well evaluated with the approximate formulas 
(\ref{ap_mW})-(\ref{ap_Zmf}), leading to
$\hat C_2 \sim \sqrt{\pi} m_W/4k\sqrt{R}$ for $W$ and 
$\hat C_4 \sim \sqrt{\pi} m_Z/4k\sqrt{(1 + \sph^2) R} $ for $Z$.  
Insertion of these gives
\bea
 \lmd_{WWH} & \simeq & \frac{g_A\sqrt{k}}{\pi R\zp}\sin\thH\cos\thH
\simeq gm_W\cdot p_{\rm H} |\cos\thH| ~, 
 \cr
 \noalign{\kern 10pt}
 \lmd_{ZZH} & \simeq & \frac{g_A\sqrt{k}(1+\sph^2)}{\pi R \zp}
 \sin\thH\cos\thH
 \simeq \frac{gm_Z}{\cos\thw}\cdot p_{\rm H} |\cos\thH| ~, 
 \label{lmds}
\eea
where $g$ is defined in (\ref{def_g}) and $p_{\rm H}\equiv\sgn(\tan\thH)$. 
\ignore{The expressions in (\ref{lmds}) are valid up to $\cO(1/k\pi R)$. }
Both $ \lmd_{WWH} $ and $ \lmd_{ZZH} $ are suppressed by a factor
$\cos \thH$  compared with the corresponding couplings in the
standard model.  Unless $\thH$ is very small, this gives substantial
suppression which can be checked in the coming experiments at LHC.
This is a generic prediction in the gauge-Higgs unification in the warped
spacetime.  It does not depend on the details of the model such as fermion 
content and couplings.

\subsection{Vanishing $WWH^{(n)}$ and $ZZH^{(n)}$ couplings}

There is a Kaluza-Klein (KK)  tower of the 4D Higgs field. 
The 4D Higgs field is a part of the fifth dimensional component of gauge
potentials.   It is associated with the Yang-Mills AB phase
along the fifth dimension, and is a physical degree of freedom.  
Its KK excited states, however,  are unphysical.  In the unitary gauge they are
eliminated to be absorbed by the KK excited states of the four-dimensional 
gauge fields.  One may wonder if these KK excited states of the 4D Higgs field,
$H^{(n)}$, have nontrivial couplings to $W$ and $Z$.  

The $WWH^{(n)}$ and $ZZH^{(n)}$ couplings are evaluated in the same manner
as the $WWH$ and $ZZH$ couplings in Sec.\ \ref{WWHcp}.  They are given by
\be
 \cL^{(4)} = \lmd_{WW{H^{(n)}}} ~ H^{(n)}W^{(0)\mu\,\dagger}W^{(0)}_\mu 
 + \half  \lmd_{ZZ{H^{(n)}}} ~ H^{(n)}Z^{(0)\mu}Z^{(0)}_\mu+\cdots, 
\ee
where
\bea
&&\hskip -1cm
\lmd_{WW{H^{(n)}}} = g_A k\int_1^{\zp}\frac{dz}{z}\; \tl{h}^{\hat 4}_{H^{(n)}}
 \brc{\tl{h}^{\hat{\pm}}_{W} 
    \der_z\brkt{\tl{h}^{\chL}_{W}    -\tl{h}^{\chR}_{W}}
 -\der_z\tl{h}^{\hat{\pm}}_{W}
    \brkt{\tl{h}^{\chL}_{W}   -\tl{h}^{\chR}_{W}}} ~, \cr
\noalign{\kern 10pt}
&&\hskip -1cm
 \lmd_{ZZ{H^{(n)}}} =g_A k\int_1^{\zp}\frac{dz}{z}\; \tl{h}^{\hat 4}_{H^{(n)}}
 \brc{\tl{h}^{\hat{3}}_{Z}
    \der_z\brkt{\tl{h}^{\nL}_{Z}   -\tl{h}^{\nR}_{Z}}
 -\der_z\tl{h}^{\hat{3}}_{Z}\brkt{\tl{h}^{\nL}_{Z}  -\tl{h}^{\nR}_{Z}}} ~. 
\label{def_WWHn}
\eea

We first recall that the spectrum $m_n^H$$=$$k \lambda_n^H$ 
for $H^{(n)}$ ($n \ge 1$) is determined by the equation
$F_{1,1} (\lambda_n^H, \lambda_n^H\zp) = 0$.  The wave function is 
given by $\tl{h}^{\hat 4}_{H^{(n)}}(z)  = C^{\hat 4}_{\vphi,n} 
z F_{0,1} (\lambda_n^H z, \lambda_n^H \zp)$ where $C^{\hat 4}_{\vphi,n}$ is
given by (\ref{scalarC3}). 
$\lmd_{WW{H^{(n)}}}$ is evaluated  as $\lmd_{WWH}$ in the previous
subsection.  A similar expression to the first line in Eq.\ (\ref{lamWWH1}) 
is obtained where $\tl{h}^{\hat 4}_{H}$  is replaced by 
$\tl{h}^{\hat 4}_{H^{(n)}}$.  Inserting the wave function, one finds
\beqn
 \lmd_{WW{H^{(n)}}} \eql
\frac{4 g_A k^2 C^{\hat 4}_{\vphi,n} C^{\hat \pm}_{W} 
(C^{\pm_{\rm L}}_{W} - C^{\pm_{\rm R}}_{W})}{\pi^2 m_W z_\pi}
\int_1^{z_\pi} dz \, z  F_{0,1} (\lambda_n^H z , \lambda_n^H z_\pi)  \cr
\noalign{\kern 10pt}
\eql  \frac{4 g_A k^2 C^{\hat 4}_{\vphi,n} C^{\hat \pm}_{W} 
(C^{\pm_{\rm L}}_{W} - C^{\pm_{\rm R}}_{W})}{\pi^2 m_W z_\pi}
~ \frac{-1}{ \lambda_n^H } ~
F_{1,1} (\lambda_n^H , \lambda_n^H z_\pi)  \cr
\noalign{\kern 10pt}
\eql 0 ~~.
\eeqn
Similarly one finds that $ \lmd_{ZZ{H^{(n)}}}=0$.  This proves that 
the $WWH^{(n)}$ and $ZZH^{(n)}$ couplings identically vanish.

\subsection{$W^4$ and $W^2 Z^2$ couplings}

The four-dimensional gauge couplings $WWW^\dagger W^\dagger$ and 
$W W^\dagger ZZ$ are evaluated from the third term in (\ref{action5}).
One finds that
\beqn
&&\hskip -1cm
 \cL^{(4)} =
 \frac{1}{2} \,  g_{WWWW}^2  
 \Big\{ W^{(0) \dagger}_\mu W^{(0) \dagger \mu}
          W^{(0)}_\nu W^{(0) \nu} - (W^{(0) \dagger}_\mu W^{(0) \mu})^2 \Big\} \cr
 \noalign{\kern 10pt}
 &&\hskip 0.3 cm       
 + g_{WWZZ}^2 \Big\{  W^{(0)}_\mu Z^{(0) \mu} W^{(0) \dagger}_\nu Z^{(0)\nu}
     - W^{(0) }_\mu W^{(0) \dagger \mu} Z^{(0)}_\nu Z^{(0) \nu} \Big\}
\eeqn
where the couplings $g_{WWWW}^2$ and $g_{WWZZ}^2$ are expressed as
overlap integrals 
\beqn
&&\hskip -1cm
g_{WWWW}^2
= g_A^2 \int_1^{z_\pi} \frac{dz}{kz} \bigg[ 
\Big\{ \big( \tl{h}^{\chL}_{W} \big)^2
 +\frac{1}{2} \big( \tl{h}^{\hat{\pm}}_{W} \big)^2 \Big\}^2 
 + \Big\{ \big( \tl{h}^{\chR}_{W} \big)^2
 +\frac{1}{2} \big( \tl{h}^{\hat{\pm}}_{W} \big)^2 \Big\}^2  \cr
\noalign{\kern 10pt} 
&&\hskip 6cm 
+ \big( \tl{h}^{\hat{\pm}}_{W} \big)^2  
\big( \tl{h}^{\chL}_{W} + \tl{h}^{\chR}_{W} \big)^2  \bigg] ~~, 
\cr
\noalign{\kern 10pt} 
&&\hskip -1cm
g_{WWZZ}^2
= g_A^2 \int_1^{z_\pi} \frac{dz}{kz} \bigg[ 
\Big\{  \tl{h}^{\chL}_{W}  \tl{h}^{\nL}_{Z} 
   + \frac{1}{2} \tl{h}^{\hat{\pm}}_{W} \tl{h}^{\hat{3}}_{Z} \Big\}^2
+ \Big\{  \tl{h}^{\chR}_{W}  \tl{h}^{\nR}_{Z} 
   + \frac{1}{2} \tl{h}^{\hat{\pm}}_{W} \tl{h}^{\hat{3}}_{Z} \Big\}^2
\cr
\noalign{\kern 10pt}
&&\hskip 3cm
+ \frac{1}{4} \Big\{ \big( \tl{h}^{\chL}_{W} + \tl{h}^{\chR}_{W} \big)
  \tl{h}^{\hat{3}}_{Z} + \tl{h}^{\hat{\pm}}_{W}
      \big(  \tl{h}^{\nL}_{Z} +  \tl{h}^{\nR}_{Z} \big) \Big\}^2  \bigg] ~~.
\label{g_WWWW1}
\eeqn
Inserting (\ref{ap_Wmf}) and (\ref{ap_Zmf}) into (\ref{g_WWWW1}), one finds
\beqn
&&\hskip -1cm
g_{WWWW}^2 \simeq \frac{g_A^2}{\pi R} = g^2 ~~, \cr
\noalign{\kern 10pt}
&&\hskip -1cm 
g_{WWZZ}^2 \simeq  \frac{g_A^2}{\pi R (1+ \sph^2 )} 
    = g^2 \cos^2 \theta_W  ~~,
\label{g_WWWW2}
\eeqn
which coincide with couplings in the standard model.  

The quartic gauge couplings are approximately independent of $\thH$
in the gauge-Higgs unification in the warped spacetime, while  in flat spacetime 
they deviate from the values in the standard model.    The approximate
universality of the gauge couplings in the warped spacetime is guaranteed 
by the approximately uniform distribution of the wave functions of 
the gauge bosons in the entire fifth dimension 
except for the tiny vicinity of the TeV brane.

\subsection{$WWHH$ and $ZZHH$ couplings}

The four-dimensional gauge-Higgs couplings $WW^\dagger HH$ and 
$ZZ HH$ are evaluated from the fourth term in (\ref{action5}).
One finds that
\beeq
 \cL^{(4)} =
 -\frac{1}{4} \,  \lambda_{WWHH}^2  
 W^{(0) \dagger}_\mu W^{(0) \mu}  H^{(0)} H^{(0)}
 - \frac{1}{8} \,  \lambda_{ZZHH}^2  
 Z^{(0)}_\mu Z^{(0) \mu}  H^{(0)} H^{(0)}
 \label{action_WWHH}
\eneq
where the couplings $\lambda_{WWHH}^2$ and $\lambda_{ZZHH}^2$ 
are expressed as overlap integrals 
\beqn
&&\hskip -1cm
\lambda_{WWHH}^2
=  k g_A^2 \int_1^{z_\pi} \frac{dz}{z} ~
\big(  \tl{h}^{\hat 4}_{H} \big)^2
\Big\{ \big( \tl{h}^{\chL}_{W} - \tl{h}^{\chR}_{W} \big)^2
 + 2 \big( \tl{h}^{\hat{\pm}}_{W} \big)^2 \Big\}   ~~,  \cr
\noalign{\kern 15pt} 
&&\hskip -1cm
\lambda_{ZZHH}^2
= k g_A^2 \int_1^{z_\pi} \frac{dz}{z}  ~
\big(  \tl{h}^{\hat 4}_{H} \big)^2
\Big\{ \big( \tl{h}^{\nL}_{Z} -  \tl{h}^{\nR}_{Z} \big)^2
 + 2 \big(  \tl{h}^{\hat{3}}_{Z} \big)^2 \Big\}   ~~.  
\label{lam_WWHH1}
\eeqn

As in the case of the cubic couplings $\lambda_{WWH}$ and $\lambda_{ZZH}$,
the Higgs wave function $\tl{h}^{\hat 4}_{\vph,0}$ is localized near the 
TeV brane so that the overlap integrals suffer from nontivial $\thH$
dependence.  With (\ref{ap_Wmf}) and (\ref{ap_Zmf}) inserted, 
the integrals in (\ref{lam_WWHH1}) are evaluated to be 
\beqn
&&\hskip -1cm
\lambda_{WWHH}^2
\simeq \frac{g_A^2}{\pi R} \Big( 1 - \frac{2}{3} \sin^2 \thH \Big)
= g^2 \Big( 1 - \frac{2}{3} \sin^2 \thH \Big) ~~, \cr
\noalign{\kern 10pt}
&&\hskip -1cm
\lambda_{ZZHH}^2
\simeq \frac{g_A^2 (1 + \sph^2 )}{\pi R} 
\Big( 1 - \frac{2}{3} \sin^2 \thH \Big)
\simeq \frac{g^2}{\cos^2 \theta_W}
\Big( 1 - \frac{2}{3} \sin^2 \thH \Big) ~~.
\label{lam_WWHH2}
\eeqn
Compared with the values in the standard model, these couplings are
suppressed by a factor $(1 - \frac{2}{3} \sin^2 \thH)$.

One comment is in order. The couplings defined in (\ref{lam_WWHH1}) and
(\ref{lam_WWHH2}) are to be called as the bare $WW^\dagger HH$ and 
$ZZ HH$ couplings.  In the effective theory at low energies where all heavy
modes are integrated out,   the effective $WW^\dagger HH$ and 
$ZZ HH$ couplings contain contributions coming from tree diagrams 
involving $W^{(n)}$ and $Z^{(n)}$ as intermediate states.
Their contributions may not be negligible, and need careful examination.

The suppression of the bare Higgs couplings is a generic feature of the 
gauge-Higgs unification, and should be used for testing  the scenario
by experiments.

\section{Gauge  couplings of fermions}

Inserting (\ref{gauge_md_ex}) and (\ref{fm_md_ex2}) into 
\bea
 S_{\rm gc}  \eql \int\dr^4x  \int_1^{\zp}\frac{dz}{k}
 \brc{g_A\bar{\tl{\Psi}}\gm^\mu \tl{A}_\mu\tl{\Psi}
 +\frac{g_B}{2}\bar{\tl{\Psi}}\gm^\mu B_\mu\cQ_{\rm B-L}\tl{\Psi}}  ~,
 \label{relevant_Sgc}
\eea
we obtain 
\bea
 \cL^{(4)}_{\rm gc} \eql 
 \sum_n W^{(n)}_\mu\brc{\frac{g^{W(n)}_L}{\sqrt{2}}
 \bar{\psi}^{(0)}_{L2}\gm^\mu\psi^{(0)}_{L1}
 +\frac{g^{W(n)}_R}{\sqrt{2}}\bar{\psi}^{(0)}_{R2}\gm^\mu 
 \psi^{(0)}_{R1}+\hc} \nonumber\\
 &&+\sum_n Z^{(n)}_\mu \sum_{i=1}^2\brc{
 g^{Z(n)}_{Li}\bar{\psi}^{(0)}_{Li}\gm^\mu\psi^{(0)}_{Li}
 +g^{Z(n)}_{Ri}\bar{\psi}^{(0)}_{Ri}\gm^\mu\psi^{(0)}_{Ri}}
 \nonumber\\
 &&+\sum_n A^{\gm(n)}_\mu \sum_{i=1}^2
 \brc{g^{\gm(n)}_{Li}\bar{\psi}^{(0)}_{Li}\gm^\mu\psi^{(0)}_{Li}
 +g^{\gm(n)}_{Ri}\bar{\psi}^{(0)}_{Ri}\gm^\mu\psi^{(0)}_{Ri}}
 +\cdots, 
\eea
where the ellipsis denotes terms involving the massive K.K. modes of 
the fermions. 
The 4D gauge couplings are given by
\bea 
 g^{W(n)}_L 
 \defa g_A\int_1^{\zp}\frac{dz}{k}\;
 \brc{ | \tl{f}^q_{L,0} |^2 \tl{h}^{\chL}_{W^{(n)}}
 + | \tl{f}^Q_{L,0} |^2 \tl{h}^{\chR}_{W^{(n)}}
 +\sqrt{2}\Im\brc{\brkt{\tl{f}^Q_{L,0}}^*
 \tl{h}^{\hat{\pm}}_{W^{(n)}}\tl{f}^q_{L,0}}}  ~~,  \nonumber\\
 \noalign{\kern 10pt}
 g^{\gm,Z(n)}_{Li}
 \defa (-)^{i-1}\frac{g_A}{2}\int_1^{\zp}\frac{dz}{k}
 \Big[ | \tl{f}^q_{L,0} |^2 \tl{h}^{\nL}_{\gamma, Z^{(n)}}
    + | \tl{f}^Q_{L,0} |^2  \tl{h}^{\nR}_{\gamma, Z^{(n)}}\Big. 
 \nonumber\\
 && \hspace{3cm}\Big. 
 +\sqrt{2} \Im \brc{ \brkt{\tl{f}^Q_{L,0}}^*
      \tl{h}^{\hat{3}}_{\gamma, Z^{(n)}  }      \tl{f}^q_{L,0}    }  \Big]
 \nonumber\\
 \noalign{\kern 10pt}
 &&
 +\frac{g_B q_{\rm B-L}}{2}\int_1^{\zp}\frac{dz}{k}\tl{h}^B_{\gamma, Z^{(n)}}
 \brc{ | \tl{f}^q_{L,0} |^2+ | \tl{f}^Q_{L,0} |^2} ~~,
 \label{def_gL}
\eea
where $q_{\rm B-L}$ is an eigenvalue of $\cQ_{\rm B-L}$. 
The index~$i=1,2$ denotes the $\suR$-doublet index. 
The same expressions hold for the right-handed (R) components where 
$L$ is replaced by $R$.
The wave functions  $\tl{h}^{\nL}_{\gamma^{(n)}}$, $ \tl{h}^{\nL}_{Z^{(n)}}$ etc. are
given by (\ref{zero_photon}), (\ref{nonzero_photon}) and by (\ref{neutral_Cs3}),
respectively.  In a simplified model without boundary mass terms there results
nontrivial couplings of right-handed fermions to $W$ bosons.

From the approximate expressions of the mode functions~(\ref{ap_Wmf}), 
(\ref{ap_Zmf}), and (\ref{ap_fmf1}), 
the 4D gauge couplings are found to be 
\bea
&&\hskip -1cm
 g^{W(0)}_L \simeq \frac{g_A}{\sqrt{\pi R}} = g  ~,  
 \label{gWL}   \\
\noalign{\kern 10pt}
&&\hskip -1cm
 g^{Z(0)}_L \simeq  \frac{(-1)^{i-1}g_A-g_B q_{\rm B-L}\sph\cph}
 {2\sqrt{(1+\sph^2)\pi R}}
 \simeq \frac{g}{\cos\thw}\brc{\frac{(-1)^{i-1}}{2}-q_{\rm EM}\sin^2\thw}  ~,  
  \label{gZL}   \\
\noalign{\kern 5pt}
&&\hskip -1cm
 g^{\gm(0)}_i    = eq_{\rm EM} ~,  
 \label{gEM}
\eea
where $e\equiv g_A\sin\thw/\sqrt{\pi R}=g\sin\thw$ is the $U(1)_{\rm EM}$ 
gauge coupling constant and $q_{\rm EM}\equiv \brc{(-1)^{i-1}+q_{\rm B-L}}/2$ 
is the electromagnetic charge. 
The relation (\ref{thW}) has been made use of in the second equality in (\ref{gZL}). 
Note that Eqs.(\ref{gWL}), (\ref{gZL}) and (\ref{gEM}) agree with 
the counterparts in the standard model.
Rigorously speaking, the couplings $g^{W(0)}_L$ and $g^{Z(0)}_L$
have small dependence on the  parameter $\alp$, 
which leads to tiny violation of the universality in weak
interactions as discussed in Ref.~\citen{HNSS}.   It was found that there results 
violation of the $\mu$-$e$ universality of $O(10^{-8})$, which is well in the 
experimental bound.

$g^{W(0)}_R$ and $g^{Z(0)}_R$ for  a multiplet $\Psi$, however,  substantially 
deviate from the standard model values.
For instance, one finds $g^{W(0)}_R \simeq g(1 - \cos\thH)/2$. 
This is because the mode functions of the right-handed fermions are localized 
near the TeV brane for $\alp>1$. 
Since  K.K. excited states are also  localized near the TeV brane, 
the mixing with K.K. excited states becomes strong, leading to the deviation. 
The problem can be avoided by introducing boundary fields~$\chi_{R,L}$ 
with boundary mass mixing with  bulk fermions, \ie, 
$\mu_q,\mu_Q\neq 0$ as in the action~(\ref{L_ferm}).   
It can be arranged such that right-handed components 
mainly consist of boundary fields on the Planck brane 
so that the deviation of the gauge couplings from the standard model 
become small enough.

\section{Summary}

Gauge-Higgs unification in warped spacetime has many attractive features.
It identifies the 4D Higgs field with a zero mode of the extra-dimensional 
components of the gauge fields, or the Yang-Mills AB phase in the extra 
dimensions.  The Higgs couplings are determined by the gauge
principle and the structure of background spacetime.  

\ignore{
In the Randall-Sundrum
warped spacetime the wave functions of the 4D gauge fields are almost
uniformly distributed over the entire fifth dimension except in the vicinity
of the TeV brane, whereas the wave function of the 4D Higgs field is localized
near the TeV brane.  This leads to important predictions which can be
tested experimentally.
}

Although the wave functions of the $W$ and $Z$ bosons substantially vary
as $\thH$, the $WWZ$, $WWWW$, and $WWZZ$  couplings in the warped 
spacetime remain nearly the same as in the standard model.  However, these
couplings considerably deviate from the standard model 
in the flat spacetime,
thus contradicting the LEP2 data on the $W$ pair production.  
These stem from the fact that 
the wave functions of the gauge bosons in the warped spacetime remain 
almost constants except in a tiny region near $y=\pi R$ 
even at $\thH\neq 0$ 
while they are deformed substantially to have 
nontrivial $y$-dependence in the entire region in the flat spacetime. 
The warped spacetime saves the universality of the gauge couplings.

The important deviation from the standard model shows up in the Higgs couplings.
We have shown that the $WWH$ and $ZZH$ couplings are suppressed by a factor 
$\cos \thH$ compared with those in the standard model whereas 
the bare $WWHH$ and $ZZHH$ couplings are suppressed by a factor 
$1 - \frac{2}{3} \sin^2 \thH$.  
The precise content of matter fields affects the location of the global
minimum of the effective potential  $V_\eff (\thH)$.
Once the value of $\thH$ is determined, 
the wave functions of $W$, $Z$, and $H$ are fixed as functions of
$\thH$.  Hence the suppression of the Higgs couplings to $W$ and $Z$
is a generic feature of the gauge-Higgs unification, and is independent of the
details of the model.  It can be used to test the scenario. 
A similar suppression of these Higgs couplings are recently discussed 
in Ref.~\citen{GGPR} with detailed phenomenological studies in the context of 
models where the Higgs emerges from a strongly-interacting sector 
as a pseudo-Goldstone boson, which are closely related to 
the gauge-Higgs unification models. 
The phenomenological study of our results such as detailed detectability 
at LHC/ILC is an important issue and 
need to  be investigated. 

As briefly discussed in the present paper, additional brane interactions are
necessary to have a realistic spectrum and gauge couplings of 
fermions.  Wave functions of fermions sensitively depend on such interactions,
and so do Yukawa couplings.  In Ref.~\citen{HNSS} it is shown that Yukawa 
couplings in a model without brane interactions are also suppressed 
compared with those in the standard model.  It is expected that Yukawa couplings
are suppressed  in a more realistic model with brane interactions as well, but
definitive statements must be awaited until precise form of the action is
specified in the fermion sector.

\vskip .5cm

\section*{Acknowledgements}
This work was supported in part by JSPS fellowship No.\ 0509241 (Y.S.), 
and by  Scientific Grants from the Ministry of 
Education and Science, Grant No.\ 17540257,
Grant No.\ 13135215,  Grant No.\ 18204024,  
and Grant No.\ 19034007(Y.H.).    One of the authors (Y.H.) would 
like to thank the Aspen Center for Physics for its hospitality where
a part of this work was done.

\appendix

\section{$SO(5)$ generators}
The spinorial representation of the $SO(5)$ generators~$T^I$ is given by 
\bea
 T^{\aL} \defa \frac{1}{2}\mtrx{\sgm_a}{}{}{0_2}, \;\;\;
 T^{\aR} \equiv \frac{1}{2}\mtrx{0_2}{}{}{\sgm_a}, \;\;\;
 T^{\hat{a}} \equiv \frac{i}{2\sqrt{2}}
 \mtrx{}{\sgm_{\hat{a}}}{-\sgm_{\hat{a}}^\dagger}{}, \label{SO_gen}
\eea
where $T^{\hat{a}}$ ($\hat{a}=1,2,3,4$) 
and $T^{\aL,\aR}$ ($\aL,\aR=1,2,3$) are respectively 
the generators of $SO(5)/SO(4)$ and $SO(4)\sim \suL\times \suR$,  
and $\sgm_{\hat{a}}\equiv (\vec{\sgm},-i1_2)$. 
They are normalized as 
\be
 \tr(T^IT^J) = \frac{1}{2}\dlt^{IJ}, 
\ee
where $I,J=(\aL, \aR, \hat{a})$.

\section{Useful formulae for Bessel functions}
Here we collect useful formulae for the Bessel functions. 
$J_\alp(z)$ and $Y_\alp(z)$ denote the Bessel functions 
of the first and second kinds, respectively. 
\bea
 J_\alp(z) \defa \brkt{\frac{z}{2}}^\alp
 \sum_{n=0}^\infty \frac{(-z^2/4)^n}{n!\Gm(\alp+n+1)}, 
 \label{def_J} \\ 
 Y_\alp(z) \defa 
 \begin{cases}
 \myfrac{1}{\sin\pi\alp}\brc{\cos\pi\alp\cdot J_\alp(z)
  -J_{-\alp}(z)} &\hbox{for } \alpha \not= \hbox{ an integer,}\cr
 \noalign{\kern 10pt}
  \myfrac{1}{\pi}\sbk{\myfrac{\der J_\alp(z)}{\der\alp}
  -(-1)^n\myfrac{\der J_{-\alp}(z)}{\der\alp}}_{\alp=n}
   &\hbox{for } \alpha=n= \hbox{ an integer.}
 \end{cases}
\eea
Their behavior for  $\abs{z}\gg 1$ is given by
\bea
 J_\alp(z) & \sim & \sqrt{\frac{2}{\pi z}}\cos\brkt{z-\frac{(2\alp+1)\pi}{4}}, 
 \nonumber\\
 Y_\alp(z) & \sim & \sqrt{\frac{2}{\pi z}}\sin\brkt{z-\frac{(2\alp+1)\pi}{4}}. 
 \label{asymp_bhv}
\eea
These Bessel functions satisfy the following relations. 
\bea
 Z_{\alp-1}(z)+Z_{\alp+1}(z) \eql \frac{2\alp}{z}Z_\alp(z), \nonumber\\
 \frac{dZ_\alp(z)}{dz} = \frac{\alp}{z}Z_\alp(z)-Z_{\alp+1}(z) 
 \eql Z_{\alp-1}(z)-\frac{\alp}{z}Z_\alp(z), \nonumber\\
 J_\alp(z)Y_{\alp-1}(z)-Y_\alp(z)J_{\alp-1}(z) \eql \frac{2}{\pi z}, 
 \label{fml_JY}
\eea
\bea
 \int^z\dr z\;zZ_\alp(\lmd z)\tl{Z}_\alp(\lmd z) \eql 
 \frac{z^2}{4}\left\{2Z_\alp(\lmd z)\tl{Z}_\alp(\lmd z)
 -Z_{\alp-1}(\lmd z)\tl{Z}_{\alp+1}(\lmd z) \right.\nonumber\\
 &&\hspace{1cm} \left. 
 -Z_{\alp+1}(\lmd z)\tl{Z}_{\alp-1}(\lmd z)\right\}, 
 \label{int_fml_Z}
\eea
where $Z_\alp(z)$, $\tl{Z}_\alp(z)$ are linear combinations of 
$J_\alp(z)$ and $Y_\alp(z)$.

\section{Properties of $F_{\alp,\bt}(u,v)$}
Using the Bessel functions, we define a function 
\be
 F_{\alp,\bt}(u,v) \equiv J_\alp(u)Y_\bt(v)-Y_\alp(u)J_\bt(v). 
 \label{def_Fs}
\ee
Using (\ref{fml_JY}), this satisfies the following relations, 
\bea
 F_{\alp,\bt}(u,v) \eql -F_{\bt,\alp}(v,u), \nonumber\\
 F_{\alp,\alp-1}(u,u) \eql -F_{\alp,\alp+1}(u,u) = \frac{2}{\pi u}, 
 \nonumber\\
 F_{\alp+1,\bt}(u,v)+F_{\alp-1,\bt}(u,v) \eql 
 \frac{2\alp}{u}F_{\alp,\bt}(u,v), \nonumber\\
 F_{\alp-1,\alp}(u,v)F_{\alp,\alp-1}(u,v) 
 \eql F_{\alp-1,\alp-1}(u,v)F_{\alp,\alp}(u,v)-\frac{4}{\pi^2uv}. 
 \label{propty_Fs}
\eea
For non-integer~$\alp$, we can express $F_{\alp,\alp}(u,v)$, 
$F_{\alp,\alp-1}(u,v)$, and $F_{\alp-1,\alp}(u,v)$ solely 
in terms of the Bessel function of the first kind as 
\bea
 F_{\alp,\alp}(u,v) \eql \frac{1}{\sin\pi\alp}
 \brc{J_{-\alp}(u)J_\alp(v)-J_\alp(u)J_{-\alp}(v)}, \nonumber\\
 F_{\alp,\alp-1}(u,v) \eql \frac{1}{\sin\pi\alp}\brc{
 J_\alp(u)J_{1-\alp}(v)+J_{-\alp}(u)J_{\alp-1}(v)}, \nonumber\\
 F_{\alp-1,\alp}(u,v) \eql -\frac{1}{\sin\pi\alp}\brc{
 J_{\alp-1}(u)J_{-\alp}(v)+J_{1-\alp}(u)J_\alp(v)}. 
\eea
Using these expressions and (\ref{def_J}), we obtain 
for $z\geq 1$ and $\lmd z \ll 1$, 
\bea
 F_{\alp,\alp}(\lmd,\lmd z) \eql \frac{z^\alp-z^{-\alp}}{\pi\alp}
 \brc{1+\cO(\lmd^2z^2)}, \nonumber\\
 F_{\alp-1,\alp}(\lmd,\lmd z) \eql -\frac{2}{\pi\lmd z^\alp}
 -\frac{\lmd}{2\pi\alp(1-\alp)}\brc{z^\alp-(1-\alp)z^{-\alp}
 -\alp z^{2-\alp}}\brc{1+\cO(\lmd^2 z^2)} \nonumber\\
 \eql -\frac{2}{\pi\lmd z^\alp}\brc{1+\cO(\lmd^2z^2,\lmd^2z^{2\alp})}, 
 \nonumber\\
 F_{\alp,\alp-1}(\lmd,\lmd z) \eql \frac{2}{\pi\lmd z^{1-\alp}} \nonumber\\
 &&+\frac{\lmd}{2\pi\alp(1-\alp)}\brc{z^{1-\alp}-\alp z^{\alp-1}
 -(1-\alp)z^{\alp+1}}\brc{1+\cO(\lmd^2 z^2)} \nonumber\\ 
 \eql \frac{2}{\pi\lmd z^{1-\alp}}\brc{1
 +\cO(\lmd^2z^2,\lmd^2z^{2(1-\alp)})}. 
 \label{asymp_F}
\eea
From (\ref{int_fml_Z}), we obtain 
\bea
 \int_1^{\zp}dz\;zF_{\alp,\bt}^2(\lmd_1 z,\lmd_2) 
 \eql \sbk{\frac{z^2}{2}\brc{F_{\alp,\bt}^2(\lmd_1 z,\lmd_2) 
 -F_{\alp+1,\bt}(\lmd_1 z,\lmd_2)F_{\alp-1,\bt}(\lmd_1 z,\lmd_2)
 }}_1^{\zp} \nonumber\\
 \eql \left[\frac{z^2}{2}\left\{F_{\alp,\bt}^2(\lmd_1 z,\lmd_2)
 +F_{\alp\pm 1,\bt}^2(\lmd_1 z,\lmd_2) \right.\right. \nonumber\\
 &&\hspace{10mm}\left.\left. -\frac{2\alp}{\lmd_1 z}
 F_{\alp,\bt}(\lmd_1 z,\lmd_2)F_{\alp\pm 1,\bt}(\lmd_1 z,\lmd_2)
 \right\}\right]_1^{\zp}. 
\eea
In the second equality, we used the third equation of 
(\ref{propty_Fs}). 
Using this, we obtain the following integral formulae, 
which are useful for determining normalization factors 
of the mode functions.  
\ignore{
\bea
 \int_1^{\zp}\frac{dz}{k}zF_{\alp,\bt}^2(\lmd,\lmd z) 
 \eql \frac{\zp^2}{2k}\brc{F_{\alp,\bt}^2+F_{\alp-1,\bt}^2
 -\frac{2\alp}{\lmd\zp}F_{\alp,\bt}F_{\alp-1,\bt}
 -\frac{4}{\pi^2\lmd^2\zp^2}}, \nonumber\\
 \int_1^{\zp}\frac{dz}{k}zF_{\alp,\bt}^2(\lmd,\lmd z) 
 \eql \frac{\zp^2}{2k}\brc{F_{\alp,\bt}^2+F_{\alp-1,\bt}^2
 -\frac{2(\alp-1)}{\lmd\zp}F_{\alp,\bt}F_{\alp-1,\bt}
 -\frac{4}{\pi^2\lmd^2\zp^2}},
 \eea
and }
\bea
 \int_1^{\zp}\dr z\;zF^2_{\alp,\bt}(\lmd z,\lmd\zp)
 \eql \frac{1}{2}\brc{\frac{4}{\pi^2\lmd^2}+\frac{2\alp}{\lmd}
 F_{\alp,\bt}F_{\alp-1,\bt}-F^2_{\alp,\bt}-F^2_{\alp-1,\bt}}, 
 \nonumber\\
 \int_1^{\zp}\dr z\;zF^2_{\alp-1,\bt}(\lmd z,\lmd\zp)
 \eql \frac{1}{2}\brc{\frac{4}{\pi^2\lmd^2}
 +\frac{2(\alp-1)}{\lmd}F_{\alp,\bt}F_{\alp-1,\bt}
 -F_{\alp,\bt}^2-F_{\alp-1,\bt}^2}. \nonumber\\
\eea
$F_{\alp,\bt}$ in the right-hand sides are understood as 
$F_{\alp,\bt}(\lmd,\lmd\zp)$.

\end{document}